\documentclass[twocolumn,compsoc,journal]{IEEEtran}
\usepackage[T1]{fontenc}
\usepackage{array}
\usepackage{booktabs}
\usepackage{calc}
\usepackage{url}
\usepackage{multirow}
\usepackage{amsthm}
\usepackage{amsmath}
\usepackage{graphicx}
\usepackage[bookmarks=false]{hyperref}

\makeatletter

\providecommand{\tabularnewline}{\\}
\newcommand{\lyxdot}{.}

 \let\oldforeign@language\foreign@language
 \DeclareRobustCommand{\foreign@language}[1]{%
   \lowercase{\oldforeign@language{#1}}}
\theoremstyle{plain}
\newtheorem{thm}{\protect\theoremname}
\theoremstyle{definition}
\newtheorem{defn}[thm]{\protect\definitionname}
\theoremstyle{plain}
\newtheorem{prop}[thm]{\protect\propositionname}

\usepackage[caption=false,font=normalsize,labelfont=sf,textfont=sf]{subfig}

\makeatother

\providecommand{\definitionname}{Definition}
\providecommand{\propositionname}{Proposition}
\providecommand{\theoremname}{Theorem}

\begin{document}

\title{Outlier Edge Detection Using Random Graph Generation Models and Applications}

\author{Honglei~Zhang,~\IEEEmembership{Member,~IEEE,} Serkan~Kiranyaz,~\IEEEmembership{Senior Member,~IEEE,}
and~Moncef~Gabbouj,~\IEEEmembership{Fellow,~IEEE}\IEEEcompsocitemizethanks{\IEEEcompsocthanksitem Honglei~Zhang and Moncef~Gabbouj are with
the Department of Signal Processing, Tampere University of Technology,
Tampere, Finland\protect \\
E-mail: \href{mailto:honglei.zhang@tut.fi}{honglei.zhang@tut.fi},
\href{mailto:moncef.gabbouj@tut.fi}{moncef.gabbouj@tut.fi}.

\IEEEcompsocthanksitem Serkan~Kiranyaz is with the Electrical Engineering
Department, College of Engineering, Qatar University, Qatar, e-mail:
\href{mailto:mkiranyaz@qu.edu.qa}{mkiranyaz@qu.edu.qa}.

}\thanks{}}

\markboth{}{Honglei Zhang \MakeLowercase{\textit{et al.}}: Outlier Edge Detection
Using Random Graph Generation Models}

\IEEEtitleabstractindextext{
\begin{abstract}
Outliers are samples that are generated by different mechanisms from
other normal data samples. Graphs, in particular social network graphs,
may contain nodes and edges that are made by scammers, malicious programs
or mistakenly by normal users. Detecting outlier nodes and edges is
important for data mining and graph analytics. However, previous research
in the field has merely focused on detecting outlier nodes. In this
article, we study the properties of edges and propose outlier edge
detection algorithms using two random graph generation models. We
found that the edge-ego-network, which can be defined as the induced
graph that contains two end nodes of an edge, their neighboring nodes
and the edges that link these nodes, contains critical information
to detect outlier edges. We evaluated the proposed algorithms by injecting
outlier edges into some real-world graph data. Experiment results
show that the proposed algorithms can effectively detect outlier edges.
In particular, the algorithm based on the Preferential Attachment
Random Graph Generation model consistently gives good performance
regardless of the test graph data. Further more, the proposed algorithms
are not limited in the area of outlier edge detection. We demonstrate
three different applications that benefit from the proposed algorithms:
1) a preprocessing tool that improves the performance of graph clustering
algorithms; 2) an outlier node detection algorithm; and 3) a novel
noisy data clustering algorithm. These applications show the great
potential of the proposed outlier edge detection techniques.\end{abstract}


}

\maketitle

\IEEEdisplaynontitleabstractindextext{}

\IEEEpeerreviewmaketitle{}

\section{Introduction}

\IEEEPARstart{G}{raphs} are an important data representation, which
have been extensively used in many scientific fields such as data
mining, bioinformatics, multimedia content retrieval and computer
vision. For several hundred years, scientists have been enthusiastic
about graph theory and its applications \cite{newman_networks:_2010}.
Since the revolution of the computer technologies and the Internet,
graph data have become more and more important because many of the
``big'' data are naturally formed in a graph structure or can be
transformed into graphs. 

Outliers almost always happen in real-world graphs. Outliers in a
graph can be outlier nodes or outlier edges. For example, outlier
nodes in a social network graph may include: scammers who steal users'
personal information; fake accounts that manipulate the reputation
management system; or spammers who send free and mostly false advertisements
\cite{jiang_catchsync:_2014,beutel_copycatch:_2013}. Researchers
have been working on algorithms to detect these malicious outlier
nodes in graphs \cite{akoglu_graph_2014,noble_graph-based_2003,dai_detecting_2012,henderson_rolx:_2012}.
Outlier edges are also common in graphs. They can be edges that are
generated by outlier nodes, or unintentional links made by normal
users or the system. Outlier edges are not only harmful but also greatly
increase the system complexity and degrade the performance of graph
mining algorithms. In this paper, we will show that the performance
of the community detection algorithms can be greatly improved when
a small amount of outlier edges are removed. Outlier edge detection
can also help evaluate and monitor the behavior of end users and further
identify the malicious entities. However, in contrast to the focus
on the outlier node detection, there have been very few studies on
outlier edge detection. 

In this paper, we present novel outlier edge detection algorithms.
Our proposed algorithms use the clustering property of social network
graphs to detect outlier edges. The outlier score of an edge is determined
by the difference of the actual number of edges and the expected number
of edges that link the two groups of nodes that are around the edge.
We use random graph generation models to predict the number of edges
between the two groups of nodes. We evaluated the proposed algorithms
using injected edges in real-world graph data. 

Further more, we show the great potentials of the outlier edge detection
technique in the areas of graph mining and pattern recognition. We
demonstrate three different applications that are based on the proposed
algorithms: 1) a preprocessing tool for graph clustering algorithms;
2) an outlier node detection algorithm; 3) a novel noisy data clustering
algorithm. 

The rest of the paper is organized as follows: the prior art is reviewed
in Section 2; the methodology to detect outlier edges is in Section
3; evaluation of the proposed algorithms are given in Section 4; various
applications that use or benefit from outlier edge detection algorithms
are presented in Section 5; and finally, conclusions and future directions
are included in Section 6.

\section{\label{sec:Previous-Work}Previous Work}

Outliers are data instances that are markedly different from the rest
of the data \cite{hodge_survey_2004}. Outliers are often located
outside (mostly far way) from the normal data points when presented
in an appropriate feature space. It is also commonly assumed that
the number of outliers is much less than the number of normal data
points. 

Outlier detection in graph data includes outlier node detection and
outlier edge detection. Noble and Cook studied substructures of graphs
and used the Minimum Description Length technique to detect unusual
patterns in a graph \cite{noble_graph-based_2003}. Xu \emph{et al}.
considered nodes that marginally connect to a structure (or community)
as outliers \cite{xu_scan:_2007}. They used a searching strategy
to group the nodes that share many common neighbors into communities.
The nodes that are not tightly connected to any community are classified
as outliers. Gao \emph{et al}. also studied the roles of the nodes
in communities \cite{gao_community_2010}. Nodes in a community tend
to have similar attributes. Using the Hidden Markov Random Field technique
as a generative model, they were able to detect the nodes that are
abnormal in their community. Akoglu \emph{et al}. detected outlier
nodes using the near-cliques and stars, heavy vicinities and dominant
heavy links properties of the ego-network--the induced network formed
by a focal node and its direct neighbors \cite{akoglu_oddball:_2010}.
They observed that some pairs of the features of normal nodes follow
a power law and defined an outlier score function that measures the
deviation of a node from the normal patterns. Dai \emph{et al}. detected
outlier nodes in bipartite graphs using mutual agreements between
nodes \cite{dai_detecting_2012}. 

In contrast to proliferative research on outlier node detection, there
have been very few studies on outlier edge detection in graphs. Chakrabarti
detected outlier edges by partitioning nodes into groups using the
Minimum Description Length technique \cite{chakrabarti_autopart:_2004}.
Edges that link the nodes from different groups are considered as
outliers. These edges are also called weak links or weak ties in literature
\cite{easley_networks_2012}. Obviously this method has severe limitations.
First, one shall not classify all weak links as outliers since they
are part of the normal graph data. Second, many outlier edges do not
happen between the groups. Finally, many graphs do not contain easily
partitionable groups. 

Detection of missing edges (or link prediction) is the opposite technique
of outlier edge detection. These algorithms find missing edges between
pairs of nodes in a graph. They are critical in recommendation systems,
especially in e-commerce industry and social network service industry
\cite{lu_link_2011,barbieri_who_2014}. Such algorithms evaluate similarities
between each pair of nodes. A pair of nodes with high similarity score
is likely to be connected by an edge. One may use the similarity scores
to detect outlier edges. The edges whose two end nodes have a low
similarity score are likely to be the outlier edges. However, in practice,
these similarity scores do not give satisfactory performance if one
uses them to detect outlier edges.

\section{Methodology}

\subsection{Notation}

Let $G(V,E)$ denote a graph with a set of nodes $V$ and a set of
edges $E$. In this article, we consider undirected, unweighted graphs
that do not contain self-loops. We use lower case $a,\ b,\ c,$ etc.,
to represent nodes. Let $\overline{ab}$ denote the edge that connects
nodes $a$ and $b$. Because our graph $G$ is undirected, $\overline{ab}$
and $\overline{ba}$ represent the same edge. Let $N_{a}$ be the
set of neighboring nodes of node $a$, such that $N_{a}=\left\{ x\vert x\in V,\overline{xa}\in E\right\} $.
Let $S_{a}=N_{a}\cup\left\{ a\right\} $ (i.e. $S_{a}$ contains node
$a$ and its neighboring nodes). Let $k_{a}$ be the degree of node
$a$, so that $k_{a}=\left|N_{a}\right|$. Let $A$ be the adjacency
matrix of graph $G$. Let $n=\left|V\right|$ be the number of nodes
and $m=\left|E\right|$ be the number of edges of graph $G$.

Freeman defines the ego-network as the induced subgraph that contains
a focal node and all of its neighboring nodes together with edges
that link these nodes \cite{freeman_centered_1982}. To study the
properties of an edge, we define the edge-ego-network as follows: 
\begin{defn}
An edge-ego-network is the induced subgraph that contains the two
end nodes of an edge, all neighboring nodes of these two end nodes
and all edges that link these nodes. 

Let $G_{\overline{ab}}=G\left(V_{\overline{ab}},E_{\overline{ab}}\right)$
denote the edge-ego-network of edge $\overline{ab}$, where $V_{\overline{ab}}=S_{a}\cup S_{b}$
and $E_{\overline{ab}}=\left\{ \overline{xy}\vert x\in V_{\overline{ab}},y\in V_{\overline{ab}}\ \mathrm{and}\ \overline{xy}\in E\right\} $.
\end{defn}

\subsection{\label{sub:missing_edge_detection}Motivation}

Graphs representing real-world data, in particular social network
graphs, often exhibit the clustering property--nodes tend to form
highly dense groups in a graph \cite{watts_collective_1998}. For
example, if two people have many friends in common, they are likely
to be friends too. Therefore, it is common for social network services
to recommend new connections to a user using this clustering property
\cite{lu_link_2011}. As a consequence, social network graphs display
an even stronger clustering property compared to other graphs. New
connections to a node may be recommended from the set of neighboring
nodes with the highest number of common neighbors to the given node.
The common neighbors (CN) score of node $a$ and node $b$ is defined
as

\begin{equation}
s_{CN}=\left|N_{a}\cap N_{b}\right|.\label{eq:S_CN}
\end{equation}

\begin{figure}[htbp]
\begin{centering}
\textsf{}%
\begin{tabular}{cc}
\textsf{\includegraphics[width=2.5cm,height=2.5cm,keepaspectratio]{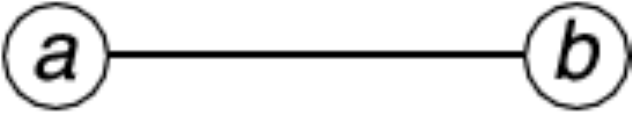}} & \includegraphics[width=4cm,height=4cm,keepaspectratio]{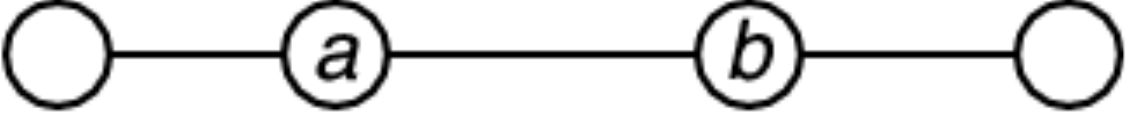}\tabularnewline
(a) & (b)\tabularnewline
 & \tabularnewline
\includegraphics[width=4cm,height=3cm,keepaspectratio]{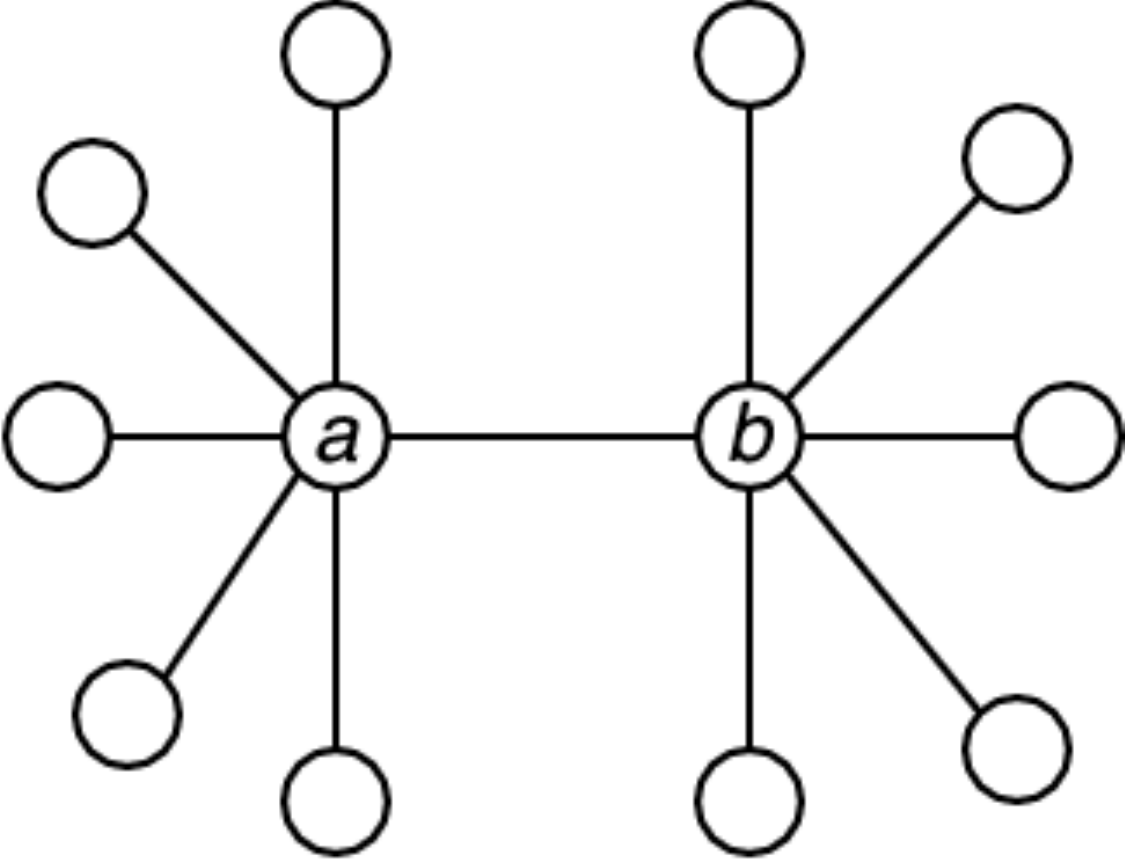} & \includegraphics[width=4cm,height=3cm]{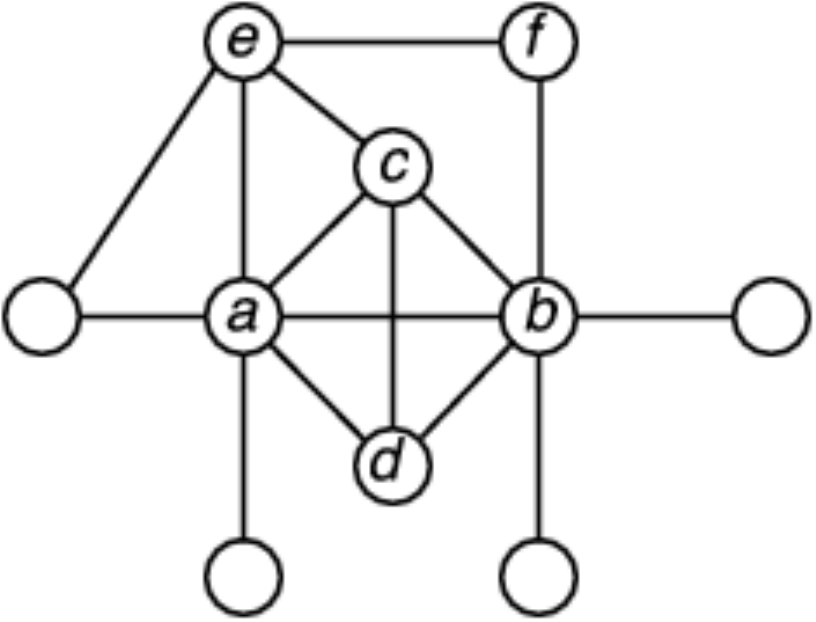}\tabularnewline
(c) & (d)\tabularnewline
\end{tabular}
\par\end{centering}

\protect\caption{\label{fig:eego_networks_cases}Different cases of edge-ego-networks
(a) $k_{a}=k_{b}=1$, $\left|N_{a}\cap N_{b}\right|=0$ (b) $k_{a}=k_{b}=2$,
$\left|N_{a}\cap N_{b}\right|=0$ (c) $k_{a}=k_{b}=6$, $\left|N_{a}\cap N_{b}\right|=0$
(d) $k_{a}=k_{b}=6$, $\left|N_{a}\cap N_{b}\right|=2$}
\end{figure}

CN score is the basis of many node similarity scores that have been
used to find missing edges \cite{lu_link_2011}. Some common similarity
indices are:
\begin{itemize}
\item Salton index or cosine similarity (Salton)

\begin{equation}
s_{Salton}=\frac{S_{CN}}{\sqrt{k_{a}k_{b}}}\label{eq:S_Salton}
\end{equation}

\item Jaccard index (Jaccard)

\begin{equation}
s_{Jaccard}=\frac{S_{CN}}{\left|N_{a}\cup N_{b}\right|}\label{eq:S_Jaccard}
\end{equation}

\item Hub promoted index (HPI)

\begin{equation}
s_{HPI}=\frac{S_{CN}}{\min(k_{a},k_{b})}\label{eq:S_HPI}
\end{equation}

\item Hub depressed index (HDI)

\begin{equation}
s_{HDI}=\frac{S_{CN}}{\max(k_{a},k_{b})}\label{eq:S_HDI}
\end{equation}

\end{itemize}
Next we shall investigate how to detect outlier edges in a social
network using the clustering property. According to this property,
if two people are friends, they are likely to have many common friends
or their friends are also friends of each other. If two people are
linked by an edge, but do not share any common friends and neither
do their friends know each other, we have good reason to suspect that
the link between them is an outlier. So, when node $a$ and node $b$
are connected by edge $\overline{ab}$, there should be edges connect
the nodes in set $S_{a}$ and the nodes in set $S_{b}$. However,
the number of connections should depend on the number of nodes in
these two groups. Let us consider the different cases as shown in
Fig. \ref{fig:eego_networks_cases}.

In these four cases, edge $\overline{ab}$ is likely to be a normal
edge in case (d) because nodes $a$ and $b$ share common neighboring
nodes $c$ and $d$, and there are connections between neighboring
nodes of $a$ and those of $b$. In the case of (a), (b) and (c),
$\left|N_{a}\cap N_{b}\right|=0$, which implies that nodes $a$ and
$b$ do not share any common neighboring nodes. However edge $\overline{ab}$
in case (c) is more likely to be an outlier edge because nodes $a$
and $b$ have each many neighboring nodes but there is no connection
between any two of these neighboring nodes. In case (a) and (b) we
do not have enough information to judge whether edge $\overline{ab}$
is an outlier edge or not. If we apply the node similarity scores
to detect outlier edges, we find that $S_{CN}=0$ for cases (a), (b)
and (c). Thus, the node similarity scores defined by Eqs. \eqref{eq:S_CN},
\eqref{eq:S_Salton}, \eqref{eq:S_Jaccard}, \eqref{eq:S_HPI} and
\eqref{eq:S_HDI} all equal to 0. For this reason, these node similarity
scores cannot effectively detect outlier edges. 

In case (c), edge $\overline{ab}$ is likely to be an outlier edge
because the expected number of edges between node $a$ together with
its neighboring nodes and node $b$ together with its neighboring
nodes is high, whereas the actual number of edges is low. So, according
to the clustering property, we propose the following definition for
the edge outlier score: 
\begin{defn}
The outlier score of an edge is defined as the difference between
the number of actual edges and the expected value of the number of
edges that link the two sets of neighboring nodes of the two end nodes
of the given edge. That is:
\begin{equation}
s_{\overline{ab}}=m_{\overline{ab}}-e_{\overline{ab}},\label{eq:outlier_score_m_e}
\end{equation}
 where $m_{\overline{ab}}$ is the actual number of edges that links
the two sets of nodes--one set is node $a$ together with its neighboring
nodes and the other set is node $b$ together with its neighboring
nodes, and $e_{\overline{ab}}$ is the expected number of edges that
link the aforementioned two sets of nodes. 
\end{defn}
We can rank the edges by their edge outlier scores defined in Eq.
\eqref{eq:outlier_score_m_e}. The edges with low scores are more
likely to be outlier edges in a graph. 

Let $\alpha\left(S,T\right)=\left|\overline{ab}\vert a\in S,b\in T\ \mathrm{and}\ \overline{ab}\in E\right|$
denote the number of edges that links the nodes in sets $S$ and $T$.
We suppose the graph $G$ is generated by a random graph generation
model. Let $\epsilon\left(S,T\right)$ denote the expected value of
the number of edges that links the nodes in sets $S$ and $T$ by
the generation model. Section \ref{sub:Expected-Number-of} describes
two generation models and the functions of calculating $\epsilon\left(S,T\right)$.
Obviously $\alpha\left(S,T\right)$ and $\epsilon\left(S,T\right)$
are symmetric functions. That is:
\begin{thm}
\label{thm:alpha_epsilon_symmetric} $\alpha\left(S,T\right)=\alpha\left(T,S\right)$
and $\epsilon\left(S,T\right)=\epsilon\left(T,S\right)$.
\end{thm}
Let $P_{a,b}$ and $R_{a,b}$ be the two sets of nodes that are related
to end nodes $a$ and $b$. Node set $R_{a,b}$ depends on set $P_{a,b}$.
The actual number of edges and the expected number of edges of the
sets of nodes related to the two end nodes may vary when we switch
the end nodes $a$ and $b$. We use the following equations to calculate
$m_{\overline{ab}}$ and $e_{\overline{ab}}$:

\begin{equation}
m_{\overline{ab}}=\frac{1}{2}\left(\alpha\left(P_{a,b},R_{a,b}\right)+\alpha\left(P_{b,a},R_{b,a}\right)\right);\label{eq:average_actual_edges}
\end{equation}

\begin{equation}
e_{\overline{ab}}=\frac{1}{2}\left(\epsilon\left(P_{a,b},R_{a,b}\right)+\epsilon\left(P_{b,a},R_{b,a}\right)\right).\label{eq:average_expected_edges}
\end{equation}

\subsection{Schemes of Node Neighborhood Sets\label{sub:Schemes-of-Sets}}

For a ego-network, Coscia and Rossetti showed the importance of removing
the focal node and all edges that link to it when studying the properties
of ego-networks \cite{coscia_demon:_2012}. It is more complicate
to study the properties of an edge-ego-network since there are two
ending nodes and two sets of neighboring nodes involved. Considering
the common nodes of the neighboring nodes and the end nodes of the
edge being investigated, we now define four schemes that capture different
configurations of these two sets. 

\begin{figure}[htbp]
\begin{centering}
\includegraphics[width=8cm,height=8cm,keepaspectratio]{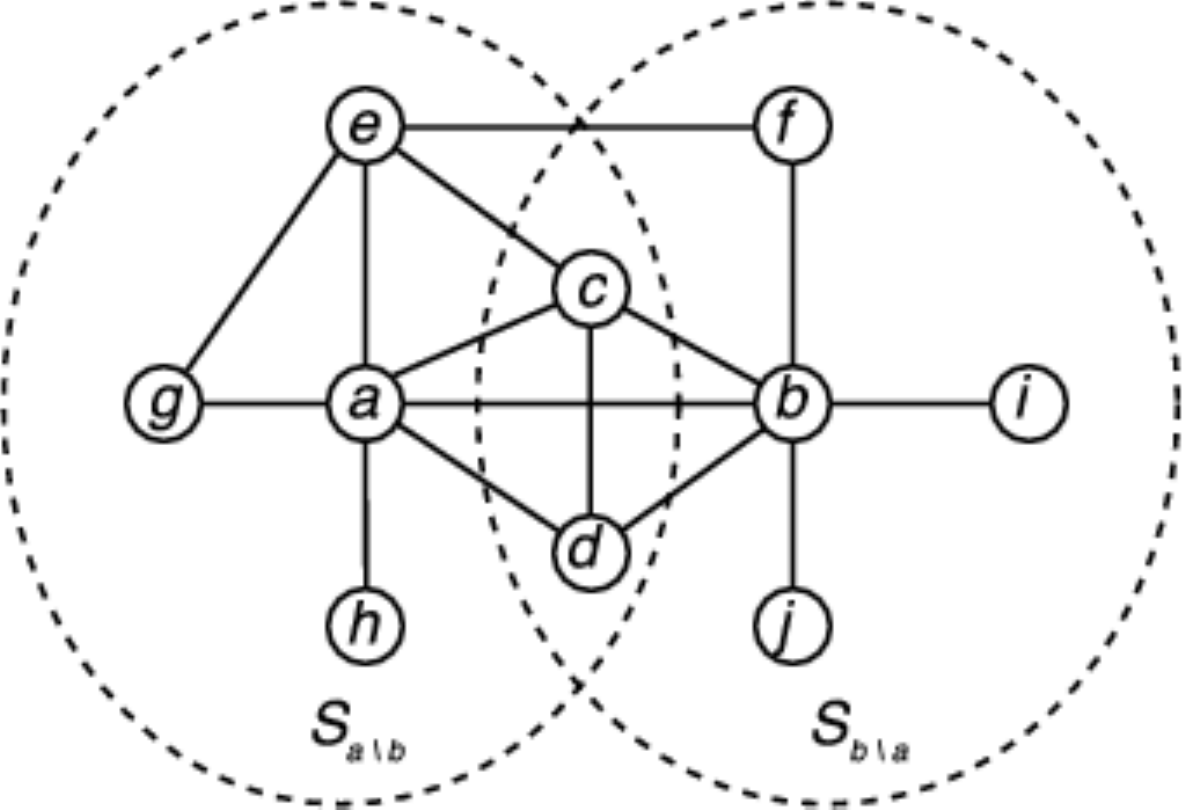}
\par\end{centering}

\protect\caption{\label{fig:eego_networks_case_d_nodes}The sets of the nodes of the
edge-ego-network $G_{\overline{ab}}$ in the case (d) of Fig. \ref{fig:eego_networks_cases}}
\end{figure}

Let $S_{a\backslash b}=S_{a}\backslash\left\{ b\right\} $ be the
set of nodes that contains node $a$ and its neighboring nodes except
node $b$. Let $N_{a\backslash b}=N_{a}\backslash\left\{ b\right\} $
be the set of nodes that contains the neighboring nodes of $a$ except
node $b$. Obviously $S_{a\backslash b}=N_{a\backslash b}\cup\left\{ a\right\} $.
Fig. \ref{fig:eego_networks_case_d_nodes} shows the edge-ego-network
$G_{\overline{ab}}$ and the two sets of nodes $S_{a\backslash b}$
and $S_{b\backslash a}$ corresponding to case (d) in Fig. \ref{fig:eego_networks_cases}.

We first define two sets of nodes that are related to node $a$ and
its neighboring nodes: $N_{a\backslash b}$ and $S_{a\backslash b}$.
Next, we define two sets of nodes that are related to node $b$ and
its neighboring nodes with regard to the sets of nodes $N_{a\backslash b}$
and $S_{a\backslash b}$: $S_{b\backslash a}\backslash S_{a\backslash b}$
and $S_{b\backslash a}$. In Fig. \ref{fig:eego_networks_case_d_nodes},
$N_{a\backslash b}=\left\{ c,d,e,g,h\right\} $, $S_{a\backslash b}=\left\{ a,c,d,e,g,h\right\} $,
$S_{b\backslash a}\backslash S_{a\backslash b}=\left\{ b,f,i,j\right\} $
and $S_{b\backslash a}=\left\{ b,c,d,f,i,j\right\} $. In the case
of a social network graph, $N_{a\backslash b}$ would consist of friends
of user (node) $a$ except $b$; $S_{a\backslash b}$ consists of
$a$ and friends of $a$ except $b$; $S_{b\backslash a}\backslash S_{a\backslash b}$
consists of $b$ and friends of $b$ except $a$ and those who are
friends of $a$; $S_{b\backslash a}$ consists of $b$ and friends
of $b$ except $a$.

Based on the set pairs of nodes $a$ and $b$, we define the following
four schemes and their meanings in the case of a social network graph.
We use superscript (1), (2), (3) and (4) to indicate the four schemes
respectively.
\begin{itemize}
\item Scheme 1 : $P_{a,b}^{(1)}=N_{a\backslash b}$ and $R_{a,b}^{(1)}=S_{b\backslash a}\backslash S_{a\backslash b}$

How many of $a$'s friends know $b$ and his friends outside of the
relationship with $a$?

\item Scheme 2 : $P_{a,b}^{(2)}=N_{a\backslash b}$ and $R_{a,b}^{(2)}=S_{b\backslash a}$

How many of $a$'s friends know $b$ and his friends?

\item Scheme 3 : $P_{a,b}^{(3)}=S_{a\backslash b}$ and $R_{a,b}^{(3)}=S_{b\backslash a}\backslash S_{a\backslash b}$

How many of $a$ and his friends know $b$ and his friends outside
of the relationship with $a$?

\item Scheme 4 : $P_{a,b}^{(4)}=S_{a\backslash b}$ and $R_{a,b}^{(4)}=S_{b\backslash a}$

How many of $a$ and his friends know $b$ and his friends?

\end{itemize}
For the edge-ego-network $G_{\overline{ab}}$ shown in Fig. \ref{fig:eego_networks_case_d_nodes},
scheme 1 examines edges $\overline{ef}$, $\overline{cb}$ and $\overline{db}$;
scheme 2 examines edges $\overline{ef}$, $\overline{ec}$, $\overline{cb}$,
$\overline{cd}$, $\overline{dc}$ and $\overline{db}$; scheme 3
examines edges $\overline{ab}$, $\overline{ef}$, $\overline{cb}$
and $\overline{db}$; scheme 4 examines edges $\overline{ab}$, $\overline{ac}$,
$\overline{ad}$, $\overline{ef}$, $\overline{ec}$, $\overline{cb}$,
$\overline{db}$, $\overline{dc}$ and $\overline{cd}$. 

Next we study the symmetric property of these four schemes. 
\begin{thm}
\label{thm:scheme2_4_symmetric}$\alpha\left(P_{a,b}^{(2)},R_{a,b}^{(2)}\right)=\alpha\left(P_{b,a}^{(2)},R_{b,a}^{(2)}\right)$
and $\alpha\left(P_{a,b}^{(4)},R_{a,b}^{(4)}\right)=\alpha\left(P_{b,a}^{(4)},R_{b,a}^{(4)}\right)$ 
\end{thm}
The proof of this theorem is given in appendix. Theorem \ref{thm:scheme2_4_symmetric}
shows that the number of edges that link the nodes from the two groups
defined in scheme 2 and scheme 4 are symmetric. That is the values
remains the same if the two end nodes are switched. We can use $m_{\overline{ab}}^{(2)}=\alpha\left(P_{a,b}^{(2)},R_{a,b}^{(2)}\right)$
and $m_{\overline{ab}}^{(4)}=\alpha\left(P_{a,b}^{(4)},R_{a,b}^{(4)}\right)$
instead of Eq. \ref{eq:average_actual_edges}.
\begin{thm}
$\epsilon\left(P_{a,b}^{(4)},R_{a,b}^{(4)}\right)=\epsilon\left(P_{b,a}^{(4)},R_{b,a}^{(4)}\right)$
\end{thm}
This theorem can be directly derived from $P_{a,b}^{(4)}=R_{b,a}^{(4)}$,
$R_{a,b}^{(4)}=P_{b,a}^{(4)}$ and Theorem \ref{thm:alpha_epsilon_symmetric}.
So $e_{\overline{ab}}=\epsilon\left(P_{a,b}^{(4)},R_{a,b}^{(4)}\right)$.
Note scheme 4 is symmetric in calculating both of the actual and expected
number of edges of the two groups.

\subsection{Expected Number of Edges Between Two Sets of Nodes\label{sub:Expected-Number-of}}

With the four schemes described above, we get the number of edges
that connect nodes from the two sets using Eq. \ref{eq:average_actual_edges}.
To calculate the outlier score of an edge by Eq. \eqref{eq:outlier_score_m_e},
we should find the expected number of edges between these two sets
of nodes. Next we will use random graph generation models to determine
the expected number of edges between these two sets of nodes.

\subsubsection{Erd\H{o}s\textendash Rényi Random Graph Generation Model}

The Erd\H{o}s\textendash Rényi model, often referred as $G(n,m)$
model, is a basic random graph generation model \cite{bollobas_random_2001}.
It generates a graph of $n$ nodes and $m$ edges by randomly connecting
two nodes by an edge and repeat this procedure until the graph contains
$m$ edges. 

Suppose we have $n$ nodes in an urn and predefined two sets of nodes
$S$ and $T$. We randomly pick two nodes from the urn. Note, the
intersection of sets $S$ and $T$ may not be empty. The probability
of picking the first node from set $S\backslash T$ is $\frac{\left|S\right|-\left|S\cap T\right|}{n}$
and the probability of picking the first node from set $S\cap T$
is $\frac{\left|S\cap T\right|}{n}$. If the first node is from set
$S$, the probability of picking the second node from set $T$ is
$\frac{\left|S\right|-\left|S\cap T\right|}{n}\frac{\left|T\right|}{n-1}+\frac{\left|S\cap T\right|}{n}\frac{\left|T\right|-1}{n-1}$.
Since the graph is undirected, we may also pick up a node from set
$T$ first and then pick up the second node from set $S$. So, the
probability that we generate an edge that connects a node set $S$
and a node from set T by randomly picking is: 
\begin{equation}
p(S,T)=\left(\left|S\right|\left|T\right|-\left|S\cap T\right|\right)\frac{2}{n(n-1)}.
\end{equation}

We repeat this procedure $m$ times to generate a graph, where $m$
is the number of edges in graph $G$. The expected number of edges
that connect the nodes in set $S$ and the nodes in set $T$ is: 
\begin{equation}
\epsilon(S,T)=\left(\left|S\right|\left|T\right|-\left|S\cap T\right|\right)\frac{2m}{n(n-1)}.\label{eq:expect_edges_ER}
\end{equation}

Note, here we ignore the duplicate edges during this procedure. This
has little impact on the final results for real-world graphs where
$m\ll n(n-1)$. In Eq. \eqref{eq:expect_edges_ER}, let 
\begin{equation}
d_{G}=\frac{2m}{n(n-1)},\label{eq:graph_density}
\end{equation}
where $d_{G}$ is the density (or fill) of graph $G$. 

Next we will find the expected number of edges under the four schemes
defined in Section \ref{sub:Schemes-of-Sets}. Since edge $\overline{ab}$
is already fixed, we should repeat the random procedure $m-1$ times.
For real-world graphs where $m\gg1$, we can safely approximate $m-1$
by $m$. 

Now we can apply Eq. \eqref{eq:expect_edges_ER} under the four schemes.
Let $k_{a}$ and $k_{b}$ be the degrees of nodes $a$ and $b$. Let
$k_{ab}=\left|N_{a}\cap N_{b}\right|$ be the number of common neighboring
nodes of nodes $a$ and $b$. The expected number of edges for each
scheme is:
\begin{itemize}
\item Scheme 1: 
\begin{equation}
e_{\overline{ab}}^{(1)}=\left(k_{a}k_{b}-\frac{1}{2}(k_{a}+k_{b})\left(1+k_{ab}\right)+k_{ab}\right)d_{G}
\end{equation}

\item Scheme 2: 
\begin{equation}
e_{\overline{ab}}^{(2)}=\left(k_{a}k_{b}-\frac{1}{2}\left(k_{a}+k_{b}\right)-k_{ab}\right)d_{G}
\end{equation}

\item Scheme 3: 
\begin{equation}
e_{\overline{ab}}^{(3)}=\left(k_{a}k_{b}-\frac{1}{2}(k_{a}+k_{b})k_{ab}\right)d_{G}
\end{equation}

\item Scheme 4: 
\begin{equation}
e_{\overline{ab}}^{(4)}=\left(k_{a}k_{b}-k_{ab}\right)d_{G}
\end{equation}

\end{itemize}

\subsubsection{Preferential Attachment Random Graph Generation Model\label{sub:Newman-Modularity-Model}}

The Erd\H{o}s\textendash Rényi model generates graphs that are lacking
some important properties of real-world data, in particular the power
law of the degree distribution \cite{newman_networks:_2010}. Next
we introduce a random graph generation model using a preferential
attachment mechanism that generates a random graph in which degrees
of each node are known. Our preferential attachment random graph generation
model (PA model) is closely related to the modularity measurement
that evaluates the community structure in a graph. Newman defines
the modularity value as the difference of the actual number of edges
and the expected number of edges of two communities \cite{newman_modularity_2006}.
The way of calculating the expected number of edges between two communities
follows preferential attachment mechanism instead of using the Erd\H{o}s\textendash Rényi
model. In the Erd\H{o}s\textendash Rényi model, each node is picked
with the same probability. However, by the preferential attachment
mechanism, the nodes with high degrees are picked with high probabilities.
Thus an edge is more likely to link nodes with a high degree. 

We can apply the preferential attachment strategy to generate a random
graph with $n$ nodes, $m$ edges and each node has a predefined degree
value. We first break each edge into two ends and put all the $2m$
ends into an urn. A node with degree $k$ will have $k$ entities
in the urn. At each round, we randomly pick two ends (one at a time
with substitution) from the urn, link them with an edge and put them
back into the urn. We repeat this procedure $m$ times. We call this
procedure Preferential Attachment Random Graph Generation model, or
PA model in short. Note, we may generate duplicate edges or even self-loops
with this procedure. Thus the expected number of edges estimated by
this model is higher than a model that does not generate duplication
edges and self-loops. This defect can be ignored when $k_{a}$ and
$k_{b}$ are small. Later we will show a method that can compensate
this bias, especially when $k_{a}$ and $k_{b}$ are large. 

If we have two nodes $a$ and $b$, the probability that an edge is
formed in each round is: 
\begin{equation}
p_{\overline{ab}}=\frac{k_{a}k_{b}}{2m^{2}}.
\end{equation}
Then the expected number of edges that link the nodes $a$ and $b$
after $m$ iterations is:
\begin{equation}
e_{\overline{ab}}=\frac{k_{a}k_{b}}{2m}.
\end{equation}

If we have two sets of nodes $S$ and $T$, the expected number of
edges that link the nodes in set $S$ and the nodes in set $T$ is:
\begin{equation}
\epsilon(S,T)=\sum_{a\in S}\sum_{b\in T}e_{\overline{ab}}=\frac{1}{2m}\sum_{a\in S}\sum_{b\in T}k_{a}k_{b}.\label{eq:expected_edges_newman}
\end{equation}

Applying Eq. \eqref{eq:expected_edges_newman} to the four schemes
defined in Section \ref{sub:Schemes-of-Sets}, we get the expected
number of edges for each scheme is
\begin{itemize}
\item Scheme 1: 
\begin{equation}
e_{\overline{ab}}^{(1)}=\frac{1}{4m}\left(\sum_{i\in P_{a,b}^{(1)}}\sum_{j\in R_{a,b}^{(1)}}k_{i}k_{j}+\sum_{i\in P_{b,a}^{(1)}}\sum_{j\in R_{b,a}^{(1)}}k_{i}k_{j}\right)\label{eq:newman_scheme_1}
\end{equation}

\item Scheme 2: 
\begin{equation}
e_{\overline{ab}}^{(2)}=\frac{1}{4m}\left(\sum_{i\in P_{a,b}^{(2)}}\sum_{j\in R_{a,b}^{(2)}}k_{i}k_{j}+\sum_{i\in P_{b,a}^{(2)}}\sum_{j\in R_{b,a}^{(2)}}k_{i}k_{j}\right)\label{eq:newman_scheme_2}
\end{equation}

\item Scheme 3: 
\begin{equation}
e_{\overline{ab}}^{(3)}=\frac{1}{4m}\left(\sum_{i\in P_{a,b}^{(3)}}\sum_{j\in R_{a,b}^{(3)}}k_{i}k_{j}+\sum_{i\in P_{b,a}^{(3)}}\sum_{j\in R_{b,a}^{(3)}}k_{i}k_{j}\right)\label{eq:newman_scheme_3}
\end{equation}

\item Scheme 4: 
\begin{equation}
e_{\overline{ab}}^{(4)}=\frac{1}{4m}\left(\sum_{i\in P_{a,b}^{(4)}}\sum_{j\in R_{a,b}^{(4)}}k_{i}k_{j}+\sum_{i\in P_{b,a}^{(4)}}\sum_{j\in R_{b,a}^{(4)}}k_{i}k_{j}\right)\label{eq:newman_scheme_4}
\end{equation}

\end{itemize}

\subsection{Edge Outlier Score Using the PA Model}

\subsubsection{Edge Outlier Score}

We may apply Eqs. \eqref{eq:newman_scheme_1}, \eqref{eq:newman_scheme_2},
\eqref{eq:newman_scheme_3} or \eqref{eq:newman_scheme_4} to Eq.
\eqref{eq:outlier_score_m_e} to calculate the outlier score of an
edge. As mentioned in Section \ref{sub:Newman-Modularity-Model},
the PA model generates graphs with duplicate edges and self-loops.
Thus the estimated expected number of edges that link two sets of
nodes are higher than an accurate model. The gap is even more significant
when the number of edges is large. To compensate for this bias, we
refine the edge outlier score function for the PA model as 
\begin{equation}
s_{\overline{ab}}=m_{\overline{ab}}^{\gamma}-e_{\overline{ab}},\label{eq:outlier_score_m2_e}
\end{equation}
where $\gamma>1$. The power function of the first term increases
the value, especially when $m_{\overline{ab}}$ is large. This eventually
compensates the bias introduced in the second term. In practice, we
normally choose $\gamma=2$.

\subsubsection{Matrix of Degree Products}

To get $e_{\overline{ab}}$ using Eqs. \eqref{eq:newman_scheme_1},
\eqref{eq:newman_scheme_2}, \eqref{eq:newman_scheme_3} or \eqref{eq:newman_scheme_4},
we should find the sum of $k_{a}k_{b}$ for every pair of nodes in
the corresponding edge-ego-network. We can store the values of $k_{a}k_{b}$
for every pair of nodes to prevent unnecessary multiplication operations
and thus reduce the processing time. However, storing this information
would require a storage space in the order of $n^{2}$, which is not
applicable when $n$ is large. We observe that we do not need to calculate
the product of the degrees for every pair of nodes in graph $G$.
What we need is the pair of nodes that appear together in every edge-ego-network. 

The distance of two nodes in a graph is defined as the length of the
shortest path between them. It is easy to see that the maximum distance
of two nodes in an edge-ego-network is 3. Next, we use the property
of the adjacency matrix to find the pairs of nodes that appear together
in edge-ego-networks. 

Let $d_{ij}$ be the distance of node $i$ and node $j$. Let $B(k)=A^{k}$,
where $A$ is the adjacency matrix of graph $G$ and $k$ is a natural
number. Let $B_{ij}(k)$ be the element of the matrix $B(k)$. Then
$B_{ij}(k)$ is the number of walks with length $k$ between node
$i$ and node $j$. If $B_{ij}(k)=0$, there is no walk with length
$k$ between nodes $i$ and $j$. 
\begin{prop}
\label{prop:power_of_a}If $d_{ij}=k$, $B_{ij}(k)\neq0$\end{prop}
\begin{IEEEproof}
If $d_{ij}=k$, there exists at least one path with length $k$ from
node $i$ to node $j$. Since a path of a graph is a walk between
two nodes without repeating nodes, there exists at least one walk
with length $k$ between the node $i$ and the node $j$. So $B_{ij}(k)\ne0$. \end{IEEEproof}
\begin{thm}
\label{thm:sum_of_power_a}Let $K(k)=B(1)+B(2)+\cdots+B(k)$. If $d_{ij}\le k$,
$K_{ij}(k)\neq0$\end{thm}
\begin{IEEEproof}
Let $d_{ij}=l$, where $l\le k$. From Proposition \ref{prop:power_of_a},
$B_{ij}(l)\ne0$. Since $B(k)$ is a nonnegative matrix where $B_{ij}(k)\ge0$,
we have $K_{ij}(k)=B_{ij}(1)+\cdots+B_{ij}(l)+\cdots+B_{ij}(k)\ne0$.
\end{IEEEproof}
According to Theorem \ref{thm:sum_of_power_a}, to find the pairs
of nodes with a distance of 3 or less, we need to find the nonzero
elements in matrix $K(3)$. Let $I$ be the indicator matrix whose
elements indicate whether the distance between a pair of nodes is
equal to or less than 3. Such that: 
\begin{equation}
I_{ij}=\begin{cases}
1 & \mathrm{if}\ K_{ij}(3)\ne0\\
0 & \mathrm{if}\ K_{ij}(3)=0
\end{cases}.
\end{equation}

Let matrix $D$ denote the degree matrix whose diagonal elements are
the degree of each node, that is: 
\begin{equation}
D_{ij}=\begin{cases}
k_{i} & \mathrm{if}\ i=j\\
0 & \mathrm{otherwise}
\end{cases}.
\end{equation}

Let 
\begin{equation}
E=\frac{1}{2m}\left(\left(DI\right)\circ\left(DI\right)^{T}\right),\label{eq:newman_expected_edge_matrix}
\end{equation}
 where $\circ$ denotes the Hadamard product of two matrices. The
value of the nonzero elements in matrix $E$ is the expected number
of edges between the two nodes under the PA model. Using matrix $E$,
we can easily calculate the edge outlier score for each scheme. For
example the outlier score of the edge $\overline{ab}$ using scheme
1 and the score function defined by Eq. \eqref{eq:outlier_score_m_e}
is: 
\begin{eqnarray}
s_{\overline{ab}}^{(1)} & = & \frac{1}{2}\left(\sum_{i\in P_{a,b}^{(1)}}\sum_{j\in R_{a,b}^{(1)}}\left(A_{ij}-E_{ij}\right)\right.\nonumber \\
 &  & +\left.\sum_{i\in P_{b,a}^{(1)}}\sum_{j\in R_{b,a}^{(1)}}\left(A_{ij}-E_{ij}\right)\right).
\end{eqnarray}

\section{Evaluation of the Proposed Algorithms}

In this section we evaluate the performance of the proposed outlier
edge detection algorithms. Due to the availability of the datasets
with identified outlier edges, we generate test data by injecting
outlier edges to real-world graphs. This experimental setup is efficient
to evaluate algorithms that detect outliers. We also evaluate the
proposed outlier detection algorithms by measuring the change of some
important graph properties when outlier edges are removed. In next
section, we will show that the proposed algorithms are not only effective
in simulated data but also powerful in solving real-world problems
in many areas.

We first inject edges to a real-world graph data by randomly picking
two nodes from the graph and linking them with an edge, if they are
not linked. The injected edges are formed randomly, and thus they
do not follow any underlying rule that generated the real-world graph.
An outlier edge detection algorithm returns the outlier score of each
edge. Given a threshold value, the edges with lower scores are classified
as outliers. 

With multiple algorithms, we vary the threshold value and record the
true positive rates and the false positive rates of each algorithm.
We use the receiver operating characteristic (ROC) curve--a plot of
true positive rates against false positive rates at various threshold
values--to subjectively compare the performance of different algorithms.
We also calculate the area under the ROC curve (AUC) value to quantitatively
evaluate the competing algorithms.

\subsection{\label{sub:Comparison_m2_e}Comparison of Different Combinations
of the Proposed Algorithm}

The proposed algorithm involves two random graph generation models
and four schemes. Two outlier score functions are proposed for the
PA Model. With the first experiment, we study the performance of different
combinations using real-world graph data. 

We take the Brightkite graph data as the test graph \cite{cho_friendship_2011}.
Brightkite is a social network service in which users share their
location information with their friends. The Brightkite graph contains
$58,228$ nodes and $214,708$ edges. The data was received from the
KONECT graph data collection \cite{kunegis_konect_2013}. 

We injected $1,000$ random ``false'' edges to the graph data. If
an algorithm yields the same outlier scores to multiple edges, we
randomly order these edges. We compare the detection results of the
algorithms using the Erd\H{o}s\textendash Rényi (ER) model and the
PA model with the combination of the four schemes explained in Section
\ref{sub:Schemes-of-Sets} and the two score functions defined in
Eqs. \eqref{eq:outlier_score_m_e} and \eqref{eq:outlier_score_m2_e}.
Table \ref{tab:Comparison_m2_e} shows the AUC values of the ROC curves
of all combinations. Bold font indicates the best score among all
of them. 

\begin{table}[htbp]
\protect\caption{\label{tab:Comparison_m2_e}AUC Values of the ROC Curves Using Brightkite
Graph Data }

\centering{}%
\begin{tabular}{ccccc}
\toprule 
\multirow{2}{*}{} & \multicolumn{2}{c}{ER Model} & \multicolumn{2}{c}{PA Model}\tabularnewline
\cmidrule{2-5} 
 & Eq. \ref{eq:outlier_score_m_e} & Eq. \ref{eq:outlier_score_m2_e} & Eq. \ref{eq:outlier_score_m_e} & Eq. \ref{eq:outlier_score_m2_e}\tabularnewline
\midrule
\addlinespace
Scheme 1 & 0.885 & 0.885 & 0.880 & 0.904\tabularnewline
\addlinespace
Scheme 2 & 0.885 & 0.885 & 0.882 & \textbf{0.905}\tabularnewline
\addlinespace
Scheme 3 & 0.878 & 0.878 & 0.873 & 0.902\tabularnewline
\addlinespace
Scheme 4 & 0.879 & 0.879 & 0.878 & 0.903\tabularnewline
\bottomrule
\addlinespace
\end{tabular}
\end{table}

From the experimental results, we see that the performance of the
PA model with score function defined by Eq. \eqref{eq:outlier_score_m2_e}
is clearly better than that of the score function defined by Eq. \eqref{eq:outlier_score_m_e}.
The term $m^{\gamma}$ in Eq. \eqref{eq:outlier_score_m2_e} increases
the value even more when $m$ is large. After the bias of the PA model
is corrected, the performance of the outlier edge detection algorithm
is greatly improved. The choice of the score function defined by Eqs.
\ref{eq:outlier_score_m_e} and \ref{eq:outlier_score_m2_e} has little
impact to the ER model based algorithms.

The results also show that the combination of the PA model and the
score function defined by Eq. \eqref{eq:outlier_score_m2_e} is superior
than other combinations by a significant margin. Scheme 2 gives better
performance than the other schemes, especially for ER Model based
algorithms. In the rest of this paper, we use scheme 2 for the ER
Model based algorithm. With the combination of the PA Model and the
score function defined by Eq. \ref{eq:outlier_score_m2_e}, the difference
between each scheme is insignificant. Because of the symmetric property
of scheme 4, we use it for the PA model with the score function defined
by Eq. \ref{eq:outlier_score_m2_e}.

\subsection{\label{sub:Comparison-of-algorithms}Comparison of Outlier Edge Detection
Algorithms}

In this section we perform comparative evaluation of the proposed
outlier edge detection algorithms against other algorithms. All test
graphs originate from the KONECT graph data collection. Table \ref{tab:Testing-Graph-Data}
shows some parameters of the test graph data. The density of a graph
is defined in Eq. \eqref{eq:graph_density}. GCC, which stands for
the global clustering coefficient, is a measure of clustering property
of a graph. It is the ratio of the number of closed triangles and
the number of connected triplet nodes. The higher GCC value is, the
stronger clustering property a graph has. 

\begin{table}[htbp]
\protect\caption{\label{tab:Testing-Graph-Data}Test Graph Data for Comparing Outlier
Edge Detection Algorithms}

\centering{}%
\begin{tabular}{cccccc}
\toprule 
 & nodes & edges & density & GCC & reference\tabularnewline
\midrule
\addlinespace
advogato & 6.5k & 51k & $1.2\times10^{-3}$ & 9.2\% & \cite{massa_bowling_2009}\tabularnewline
\addlinespace
twitter-icwsm & 465k & 835k & $3.9\times10^{-6}$ & 0.06\% & \cite{de_choudhury_how_2010}\tabularnewline
\addlinespace
brightkite & 58k & 214k & $1.3\times10^{-4}$ & 11\% & \cite{cho_friendship_2011}\tabularnewline
\addlinespace
facebook-wosn & 63k & 817k & $4.0\times10^{-4}$ & 14.8\% & \cite{viswanath_evolution_2009}\tabularnewline
\addlinespace
ca-cit-HepPh & 28k & 4.6m & $8.0\times10^{-3}$ & 28\% & \cite{leskovec_graph_2007}\tabularnewline
\addlinespace
youtube-friend & 1.1m & 3.0m & $4.6\times10^{-6}$ & 0.6\% & \cite{yang_defining_2015}\tabularnewline
\addlinespace
web-Google & 875k & 5.1m & $6.7\times10^{-6}$ & 5.5\% & \cite{leskovec_statistical_2008}\tabularnewline
\bottomrule
\addlinespace
\end{tabular}
\end{table}

We compared the performance of the two proposed algorithms (ER model
combined with scheme 2 and the score function defined by Eq. \eqref{eq:outlier_score_m_e}
and PA model combined with scheme 4 and the score function defined
by Eq. \eqref{eq:outlier_score_m2_e}) with three other algorithms
that use node similarity scores for missing edge detection. We use
the Jaccard Index and Hub Promoted Index (HPI) as defined in Eqs.
\eqref{eq:S_Jaccard} and \eqref{eq:S_HPI}. We also use the Preferential
Attachment Index (PAI) that is another missing edge detection metric
that works for outlier edge detection. The PAI for edge $\overline{ab}$
is defined as 
\begin{equation}
s_{PAI}=k_{a}k_{b}.\label{eq:s_PAI}
\end{equation}

\begin{figure}[htbp]
\begin{centering}
\textsf{\includegraphics[scale=0.9]{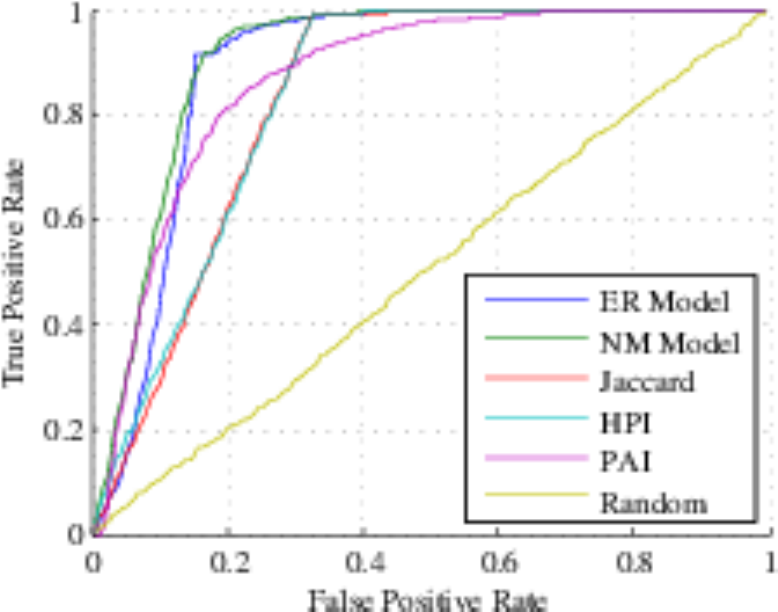}}
\par\end{centering}

\protect\caption{\label{fig:ROC-curve-algorithms}ROC curve of different algorithms
on the Brightkite graph data }
\end{figure}

Fig. \ref{fig:ROC-curve-algorithms} shows the ROC curves of different
algorithms on the Brightkite graph data. For reference, the figure
also shows an algorithm that randomly orders the edges by giving random
scores to each edge. 

As Fig. \ref{fig:ROC-curve-algorithms} shows, the ROC curve of the
algorithm that gives random scores is roughly a straight line from
the origin to the top right corner. This line indicates that the algorithm
cannot distinguish between an outlier edge and a normal edge, which
is expected. The ROC curve of an algorithm that can detect outlier
edges should be a curve above this straight line, as all algorithms
used in this experiment. As mentioned in Section \ref{sub:missing_edge_detection},
the Jaccard Index and HPI both use the number of common neighbors.
Thus their scores are all 0 for edges that connect two end nodes that
do not share any common neighbors. In real-world graphs, a large amount
of edges have a Jaccard Index or HPI value 0, especially for graphs
that contain many low degree nodes. 

The PAI value is the product of the degrees of the two end nodes of
an edge. Sorting edges with their PAI values just puts the edges with
low degree end nodes to the front. The figure shows that the PAI value
can detect outlier edges with fairly good performance. This indicates
that most of the injected edges connecting the nodes with low degrees.
Considering most of the nodes in a real-world graph are low degree
nodes, this is an expected behavior.

Fig. \ref{fig:ROC-curve-algorithms} indicates that the proposed outlier
edge detection algorithms are clearly superior to the competing algorithms.
The algorithm based on the PA model performs better than the one based
on the ER model . 

Table \ref{tab:comparison_algorithms} shows the AUC values of the
ROC curves on all test graph data. Bold font shows the best AUC values
for each test graph. 

\begin{table}[htbp]
\protect\caption{\label{tab:comparison_algorithms}AUC values of the ROC Curves on
Different Graph Data}

\centering{}%
\begin{tabular}{cccccc}
\toprule 
 & ER & PA & Jaccard & HPI & PAI\tabularnewline
\midrule
\addlinespace
advogato & 0.887 & \textbf{0.893} & 0.858 & 0.859 & 0.877\tabularnewline
\addlinespace
twitter-icwsm & 0.531 & 0.942 & 0.527 & 0.530 & \textbf{0.997}\tabularnewline
\addlinespace
brightkite & 0.885 & \textbf{0.905} & 0.833 & 0.827 & 0.873\tabularnewline
\addlinespace
facebook-wosn & 0.968 & \textbf{0.970} & 0.947 & 0.946 & 0.878\tabularnewline
\addlinespace
ca-cit-HepPh & 0.970 & 0.967 & \textbf{0.993} & 0.991 & 0.888\tabularnewline
\addlinespace
youtube-friend & 0.770 & 0.842 & 0.731 & 0.738 & \textbf{0.898}\tabularnewline
\addlinespace
web-Google & 0.985 & \textbf{0.992} & 0.944 & 0.945 & 0.859\tabularnewline
\bottomrule
\addlinespace
\end{tabular}
\end{table}

The comparison results show that the PA model algorithm gives consistently
good performance regardless of the test graph data. The experiment
also shows the correlation between the performance of the algorithms
that are based on the random graph generation model and the GCC value
of the test graph. For example, the ER model and PA model algorithms
works better on Facebook-Wosn and Brightkite graph data, which have
high GCC values as shown in Table \ref{tab:Testing-Graph-Data}. Performance
of the ER model algorithm degrades considerably on graphs with a very
low GCC value, such as the twitter-icwsm graph. This result agrees
with the fact that both the ER model and the PA model algorithms use
the clustering property of graphs. We also observe that PAI works
better on graphs with low GCC values. We estimate that these graphs
contain many star structures and two nodes with low degrees are rarely
linked by an edge. The large number of claw count (28 billion) and
small number of triangle count (38k) in twitter-icwsm graph data partially
confirm our estimation.

\subsection{Change of Graph Properties}

The proposed outlier edge detection algorithms are based on the clustering
property of graphs. Since outlier edges are defined as edges that
do not follow the clustering property, removing them should increase
the coefficients that measure this property. On the other hand, some
outlier edges (also called weak links in this aspect) serves an important
role to connect remote nodes or nodes from different communities.
Removing such edges should also extensively increase the distance
of the two end nodes. Thus the coefficients that measure the distance
between the nodes of a graph shall increase when outlier edges are
removed. In this experiment, we verify these changes caused by the
removal of the detected outlier edges. 

The global clustering coefficient (GCC) and the average local clustering
coefficient (ALCC) are the de facto measures of the clustering property
of graphs. GCC is defined in Section \ref{sub:Comparison-of-algorithms}.
Local clustering coefficient (LCC) is the ratio of the number of edges
that connect neighboring nodes of a node and the number of all possible
edges that connect these neighboring nodes. The LCC of node $a$ can
be expressed as 
\begin{equation}
c_{a}=\frac{\left|\left\{ \overline{ij}\vert i\in N_{a},j\in N_{a},\overline{ij}\in E\right\} \right|}{k_{a}\left(k_{a}-1\right)}.
\end{equation}
 ALCC is the average of the local clustering coefficients of all nodes
in the graph. 

We use diameter, the 90-percentile effective diameter (ED) and the
mean shortest path (MSP) length as distance measures between the nodes
in a graph. Diameter is the maximum shortest path length between any
two nodes in a graph. 90-percentile effective diameter is the number
of edges that are needed on average to reach 90\% of other nodes.
The mean shortest path length is the average of the shortest path
length between each pair of nodes in the graph. Note, if the graph
is not connected, we measure the diameter, ED and MSP of the largest
component in the graph.

In this experiment, we removed 5\% of the edges with the lowest outlier
score. Table \ref{tab:Graph-Properties-Changes} shows the GCC, ALCC,
Diameter, ED and MSP values before and after the outlier edges were
removed. For comparison, we also calculated values of these coefficients
after same amount of edges are randomly removed 5\% from the graph. 

\begin{table}[htbp]
\protect\caption{\label{tab:Graph-Properties-Changes}Graph Properties Changes After
Noise Edges Removal}

\centering{}%
\begin{tabular}{ccccc}
\toprule 
 & Original & ER Model & PA Model & Random\tabularnewline
\midrule
\addlinespace
GCC & 0.111 & 0.121 & 0.120 & 0.105\tabularnewline
\addlinespace
ALCC & 0.172 & 0.180 & 0.183 & 0.158\tabularnewline
\addlinespace
Diameter & 18 & 19 & 20 & 18\tabularnewline
\addlinespace
ED & 5.91 & 6.78 & 6.36 & 5.95\tabularnewline
\addlinespace
MSP & 3.92 & 4.10 & 4.10 & 3.95\tabularnewline
\bottomrule
\addlinespace
\end{tabular}
\end{table}

The results show that removing the detected outlier edges clearly
increases the GCC and ALCC values, while random edge removal slightly
decreases the values. This confirms the enhancement of the clustering
property after outlier edges are removed. The diameter, ED and MSP
values all increase when the detected outlier edges were removed.
This increase is much more significant than when random edges were
removed. This also confirms the theoretical prediction.

\section{Applications}

In this section, we demonstrate various applications that benefit
from the proposed outlier edge detection algorithms. In these applications,
we use the algorithm of the PA model combined with scheme 4 and the
score function defined by Eq. \ref{eq:outlier_score_m2_e}.

\subsection{Impact on Graph Clustering Algorithms}

Graph clustering is an important task in graph mining \cite{fortunato_community_2010,coscia_classification_2011,papadopoulos_community_2011}.
It aims to find clusters in a graph--a group of nodes in which the
number of inner links between the nodes inside the group is much higher
than that between the nodes inside the group and those outside the
group. Many techniques have been proposed to solve this problem \cite{blondel_fast_2008,danon_comparing_2005,newman_finding_2004,schaeffer_graph_2007}.

The proposed outlier edge detection algorithms are based on the graph
clustering property. They find edges that link the nodes in different
clusters. These edges are also called weak links in the literature.
With the proposed techniques, we can now remove detected outlier edges
before applying a graph clustering algorithm. This should improve
the graph clustering accuracy and reduce the computational time. 

In this application, we evaluate the performance impact of the proposed
outlier edge detection technique on different graph clustering algorithms.
We use simulated graph data with cluster structures as used in \cite{newman_finding_2004,danon_effect_2006,lancichinetti_benchmark_2008,newman_fast_2004}.
We generated test graphs of 512 nodes. The average degree of each
node is 24. The generated cluster size varies from 16 to 256. Let
$d_{out}$ be the average number of edges that link a node from the
cluster to nodes outside the cluster. Let $d$ be the average degree
of the node. Let $\mu=\frac{d_{out}}{d}$ be the parameter that indicates
the strength of the clustering structure. The smaller $\mu$ is, the
stronger the clustering structure is in the graph. We varied $\mu$
from 0.2 to 0.5. Note, when $\mu=0.5$, the graph has a very weak
clustering structure, i.e. a node inside the cluster has an equal
number of edges that link it to other nodes inside and outside the
cluster. 

We use the Normalized Mutual Information (NMI) to evaluated the accuracy
of a graph clustering algorithm. The NMI value is between 0 and 1.
The larger the NMI value is, the more accurate the graph clustering
result is. An NMI value of 1 indicates that the clustering result
matches the ground truth. More details of the NMI metric can be found
in \cite{danon_comparing_2005,ana_robust_2003}. 

We first apply graph clustering algorithms to the test graph data
and record their NMI values and computational time. Then we remove
5\% of the detected outlier edges from the test graph data, and apply
these graph clustering algorithms again to the new graph and record
their NMI values and computational time. The differences of the NMI
values and the computational time show the impact of the outlier edge
removal on the graph clustering algorithms. 

The evaluated algorithms are GN \cite{newman_finding_2004}, SLM \cite{waltman_smart_2013},
Danon \cite{danon_effect_2006}, Louvain \cite{blondel_fast_2008}
and Infomap \cite{rosvall_maps_2008}. MCL \cite{dongen_graph_2000}
is not listed since it failed to find the cluster structure from this
type of test graph data. 

We repeated the experiment 10 times and calculated the average performance.
Table \ref{tab:impact_on_graph_clustering_nmi_score} shows the NMI
values before and after outlier edges were removed. The first number
in each cell shows the NMI values of the clustering result on the
original graph and the second number shows the NMI values of the clustering
result on the graph after the outlier edges were removed. 

\begin{table}[htbp]
\protect\caption{\label{tab:impact_on_graph_clustering_nmi_score}The NMI Values Before
and After Outlier Edges Were Removed}

\centering{}%
\begin{tabular}{cccccc}
\toprule 
$\mu$ & GN & SLM & Danon & Louvain & Infomap\tabularnewline
\midrule
\addlinespace
0.2 & 0.99/1.0 & 1.0/1.0 & 0.99/1.0 & 1.0/1.0 & 1.0/1.0\tabularnewline
\addlinespace
0.25 & 0.98/0.99 & 1.0/1.0 & 0.99/0.98 & 1.0/1.0 & 1.0/1.0\tabularnewline
\addlinespace
0.3 & 0.93/0.97 & 1.0/1.0 & 0.95/0.98 & 1.0/1.0 & 0.92/1.0\tabularnewline
\addlinespace
0.35 & 0.74/0.72 & 0.96/0.94 & 0.66/0.84 & 0.90/0.86 & 0.36/0.91\tabularnewline
\addlinespace
0.4 & 0.66/0.70 & 0.83/0.81 & 0.67/0.70 & 0.84/0.81 & 0.78/0.83\tabularnewline
\addlinespace
0.45 & 0.53/0.52 & 0.71/0.67 & 0.51/0.55 & 0.68/0.60 & 0.22/0.43\tabularnewline
\addlinespace
0.5 & 0.39/0.47 & 0.58/0.56 & 0.39/0.49 & 0.51/0.53 & 0/0.47\tabularnewline
\bottomrule
\addlinespace
\end{tabular}
\end{table}

Table \ref{tab:impact_on_graph_clustering_nmi_changes} shows the
NMI value changes in percentage. A positive value indicates that the
NMI value has increased. 

\begin{table}[htbp]
\protect\caption{\label{tab:impact_on_graph_clustering_nmi_changes}Changes of Normalized
Mutual Information on Graph Clustering Algorithms in percentage}

\centering{}%
\begin{tabular}{cccccc}
\toprule 
$\mu$ & GN & SLM & Danon & Louvain & Infomap\tabularnewline
\midrule
\addlinespace
0.2 & 0.8\% & 0 & 1.0\% & 0 & 0\tabularnewline
\addlinespace
0.25 & 1.5\% & 0 & -1.0\% & 0 & 0\tabularnewline
\addlinespace
0.3 & 5.0\% & 0 & 3.5\% & 0 & 9.1\%\tabularnewline
\addlinespace
0.35 & -2.2\% & -2.1\% & 26\% & -4.9\% & 155\%\tabularnewline
\addlinespace
0.4 & 6.7\% & -2.2\% & 4.8\% & -3.0\% & 5.8\%\tabularnewline
\addlinespace
0.45 & -1.1\% & -6.2\% & 8.4\% & -12\% & 95\%\tabularnewline
\addlinespace
0.5 & 19\% & -4.4\% & 26\% & 2.4\% & $\infty$\tabularnewline
\bottomrule
\addlinespace
\end{tabular}
\end{table}

The results show that outlier edge removal improves the accuracy of
most graph clustering algorithms. The clustering accuracy of the SLM
algorithm and the Louvain algorithm decrease slightly in some cases.

Table \ref{tab:impact_on_graph_clustering_time_change} shows the
computational time changes in percentage before and after outlier
edges are removed. Negative values indicate that the computational
time is decreased. 

\begin{table}[htbp]
\protect\caption{\label{tab:impact_on_graph_clustering_time_change}Changes of Computational
Time on Graph Clustering Algorithms in percentage}

\centering{}%
\begin{tabular}{cccccc}
\toprule 
$\mu$ & GN & SLM & Danon & Louvain & Infomap\tabularnewline
\midrule
\addlinespace
0.2 & -11\% & -36\% & -3.1\% & -33\% & -47\%\tabularnewline
\addlinespace
0.25 & -18\% & 1.0\% & -1.0\% & -41\% & -16\%\tabularnewline
\addlinespace
0.3 & -9.3\% & 7.7\% & -1.4\% & -31\% & -13\%\tabularnewline
\addlinespace
0.35 & -0.3\% & -21\% & -3.5\% & -35\% & 31\%\tabularnewline
\addlinespace
0.4 & -5.7\% & -5.3\% & -3.0\% & -20\% & 17\%\tabularnewline
\addlinespace
0.45 & 2.8\% & -14.4\% & 2.1\% & -41\% & 33\%\tabularnewline
\addlinespace
0.5 & -6.7\% & -1.9\% & -3.4\% & -39\% & 55\%\tabularnewline
\bottomrule
\addlinespace
\end{tabular}
\end{table}

These results show that outlier edge removal decreases the computational
time of most algorithms used in the experiment. In some cases, SLM
and the Louvain algorithms show significant gains in computation time.
Note further that the increase of the computational time in the Infomap
algorithm leads to a crucial improvement of the clustering accuracy.

\subsection{Outlier Node Detection in Social Network Graphs}

As mentioned in Section \ref{sec:Previous-Work}, many algorithms
have been proposed to detect outlier nodes in a graph. In this section
we present a technique to detect outlier nodes using the proposed
outlier edge detection algorithm. 

In a social network service, if a user generates many links that do
not follow the clustering property, we have good reasons to suspect
that the user is a scammer. To detect this type of outlier nodes,
we can first detect outlier edges. Then we find nodes that are the
end points of these outlier edges. Nodes that are linked to many outlier
edges are likely to be outlier nodes.

In this application, we use Brightkite data for outlier node detection.
In the experiment, we rank the edges according to their outlier scores.
We take the first 1000 edges as outlier edges and rank each node according
to the number of outlier edges that it is connected to. 

Table \ref{tab:outlier_node_detection} shows the top 8 detected outlier
nodes: the node ID, the number of outlier edges that the node links,
the degree of the node, the rank of the degree among all nodes and
LCC values of the node. 

\begin{table}[htbp]
\protect\caption{\label{tab:outlier_node_detection}Outlier Node Detection Results
on Brightkite Graph}

\centering{}%
\begin{tabular}{ccccc}
\toprule 
node id & outlier edges & degree & degree rank & LCC\tabularnewline
\midrule
\addlinespace
41 & 21 & 1134 & 1 & 0.005\tabularnewline
\addlinespace
458 & 16 & 1055 & 2 & 0.001\tabularnewline
\addlinespace
115 & 9 & 838 & 4 & 0.004\tabularnewline
\addlinespace
175 & 7 & 270 & 39 & 0.001\tabularnewline
\addlinespace
989 & 7 & 270 & 40 & 0.015\tabularnewline
\addlinespace
2443 & 7 & 379 & 16 & 0.010\tabularnewline
\addlinespace
36 & 5 & 467 & 11 & 0.005\tabularnewline
\addlinespace
158 & 5 & 833 & 5 & 0.004\tabularnewline
\bottomrule
\addlinespace
\end{tabular}
\end{table}

The results show that the detected outlier nodes tend to have large
degree values. In particular, the LCC values of the detected outlier
nodes are extremely low comparing to the ALCC value (0.172) of the
graph. This shows that the neighboring nodes of the detected outlier
nodes have very weak clustering property.

\subsection{Clustering of Noisy Data}

Clustering is one of the most important tasks in machine learning
\cite{theodoridis_pattern_2008}. During the last decades, many algorithms
have been proposed, i.e. \cite{jain_data_1999,lloyd_least_1982,zhang_graph_2012}.
The task becomes more challenging when noise is present in the data.
Many algorithms, especially connectivity-based clustering algorithms,
fail over such data. In this section we present a robust clustering
algorithm that uses the proposed outlier edge detection techniques
to find correct clusters in noisy data. 

Graph algorithms have been successfully used in clustering problems
\cite{harel_clustering_2001,dong_clustering_2012}. To cluster the
data, we first build a mutual $k$-nearest neighbor (MKNN) graph \cite{brito_connectivity_1997,ozaki_using_2011}.
Let $x_{1},x_{2},\ldots,x_{n}\in R^{d}$ be the data points, where
$n$ is the number of data points and $d$ is the dimension of the
data. Let $d(x_{i},x_{j})$ be the distance between two data points
$x_{i}$ and $x_{j}$. Let $N_{k}(x_{i})$ be the set of data points
that are the $k$-nearest neighbors of the data point $x_{i}$ with
respect to the predefined distance measure $d\left(x_{i},x_{j}\right)$.
Therefore, the cardinality of the set $N_{k}(x_{i})$ is $k$. A MKNN
graph is built in the following way. The nodes in the MKNN graph are
the data points. Two nodes $x_{i}$ and $x_{j}$ are connected if
$x_{i}\in N_{k}(x_{j})$ and $x_{j}\in N_{k}(x_{i})$. The constructed
MKNN graph is unweighted and undirected. 

With a proper distance function, data points in a cluster are close
to each other whereas data points in different clusters are far away
from each other. Thus, in the constructed MKNN graph, a node is likely
to be linked to other nodes in the same cluster while the links between
the nodes in different clusters are relatively less. This indicates
that the MKNN graph has the clustering property similar to social
network graphs. 

Outlier data points are normally far away from the normal data points.
Some outlier nodes form isolated small components in the MKNN graph.
However, the outlier nodes that fall between the clusters form bridges
that connect different clusters. These bridges greatly degrade the
performance of connectivity-based clustering algorithms, such as single-linkage
clustering algorithm and complete-linkage clustering algorithm \cite{theodoridis_pattern_2008}. 

Based on these observations, we propose a hierarchical clustering
algorithm by iteratively removing edges (weak links) according to
their outlier scores. When a certain amount of outlier edges is removed,
different clusters form separate large connected components--a connected
component in a graph that contains a large proportion of the nodes,
and it is straightforward to find them in the graph. A breadth-first
search or a depth-first search algorithm can find all connected components
in a graph with the complexity of $O(n)$, where $n$ is the number
of nodes. At each iteration step, we find large connected components
in the MKNN graph and the data points that do not belong to any large
connected components are classified as outliers.

\begin{figure}[tbh]
\begin{centering}
\textsf{}%
\begin{tabular}{cc}
\textsf{\includegraphics[width=4cm]{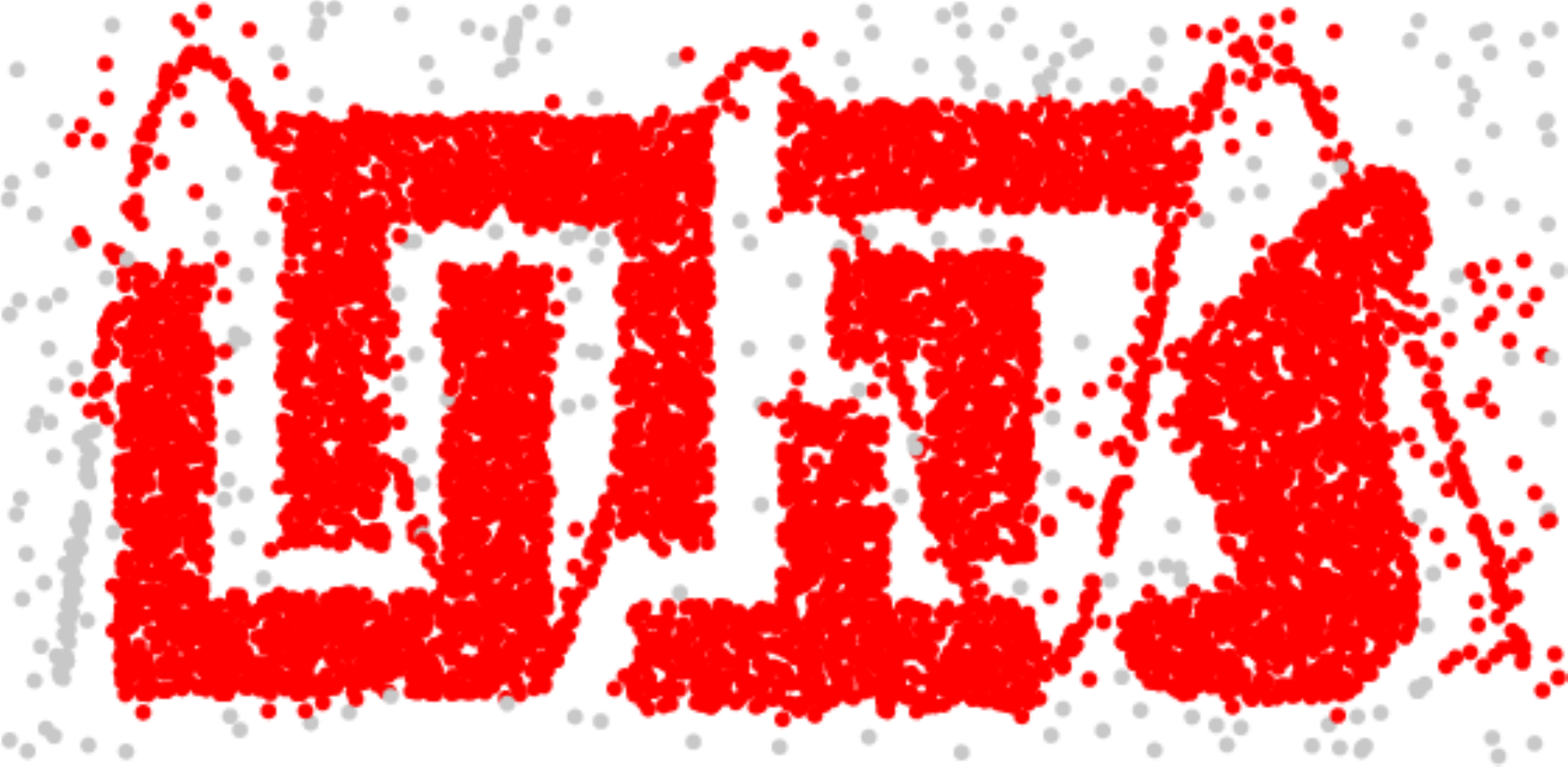}} & \includegraphics[width=4cm]{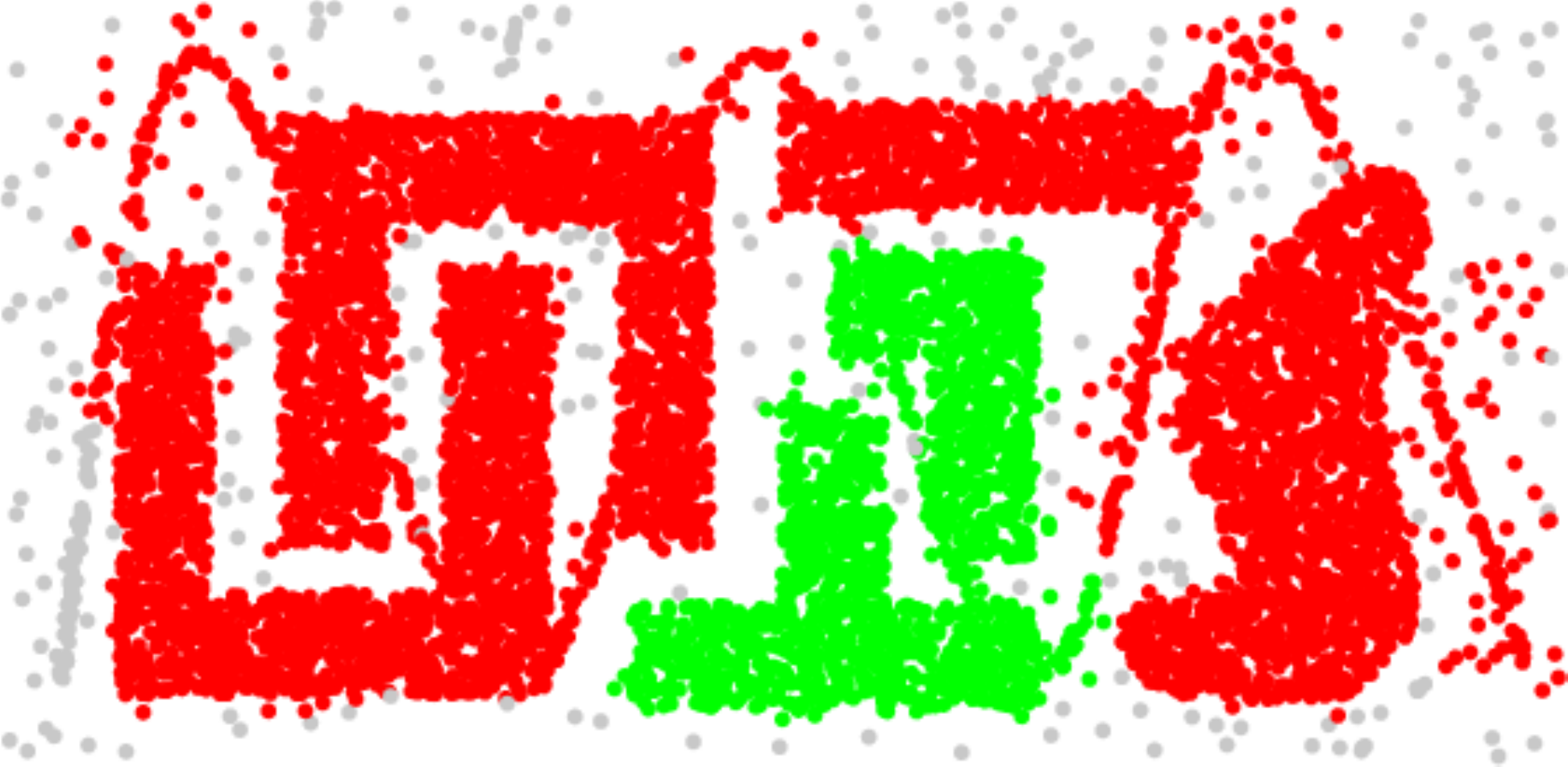}\tabularnewline
(a) & (b)\tabularnewline
 & \tabularnewline
\textsf{\includegraphics[width=4cm]{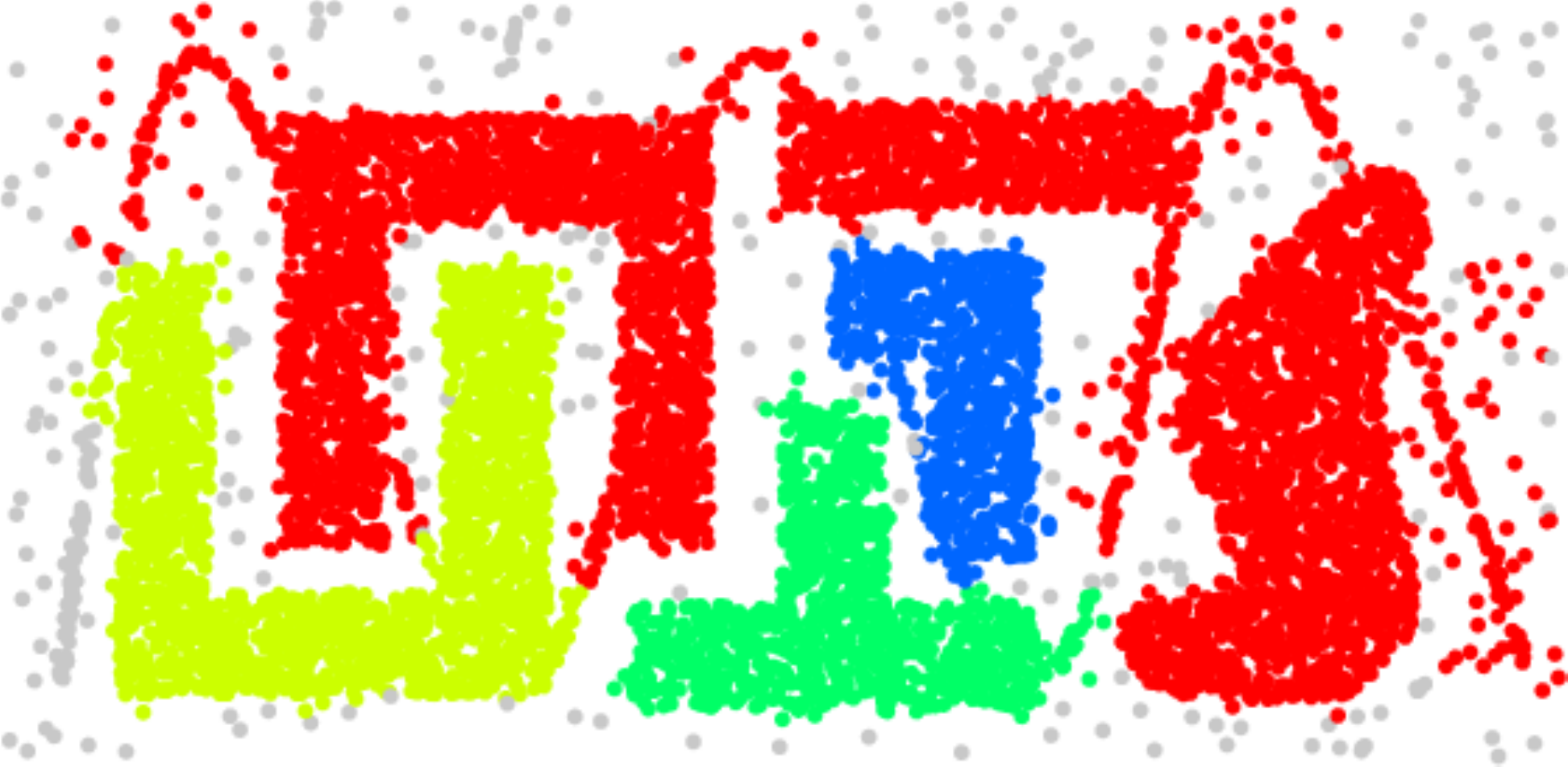}} & \textsf{\includegraphics[width=4cm]{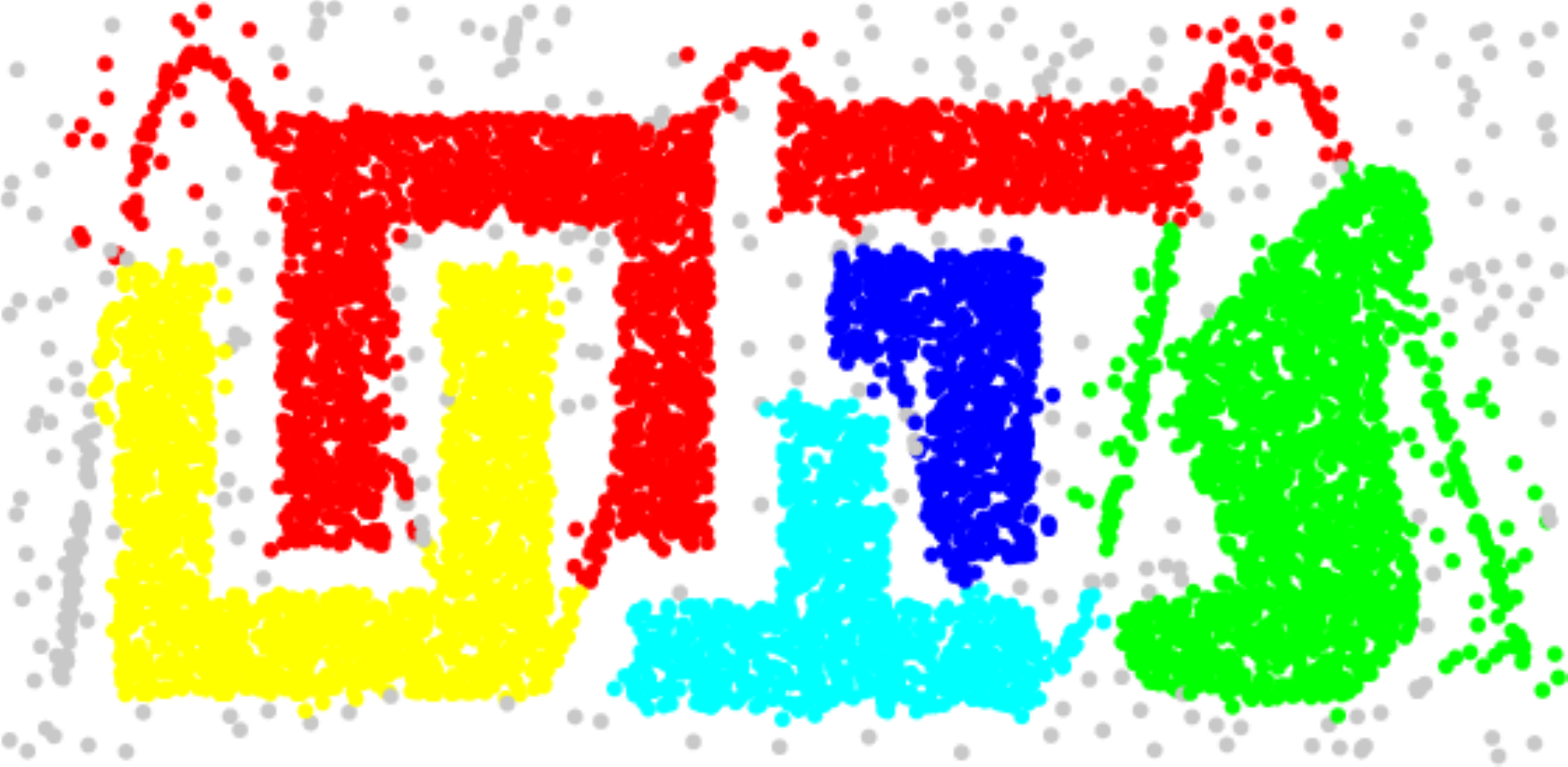}}\tabularnewline
(c) & (d)\tabularnewline
 & \tabularnewline
\textsf{\includegraphics[width=4cm]{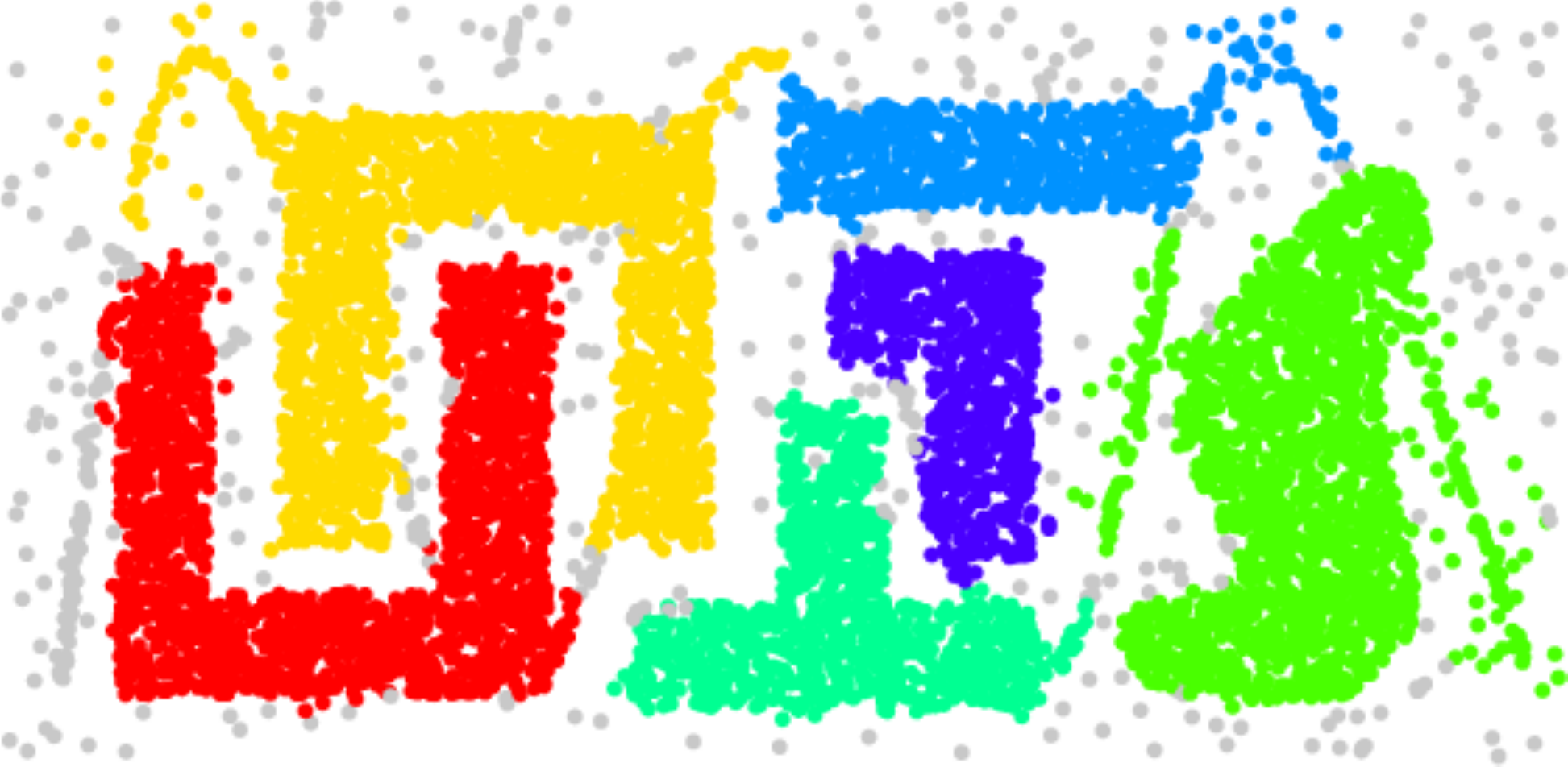}} & \textsf{\includegraphics[width=4cm]{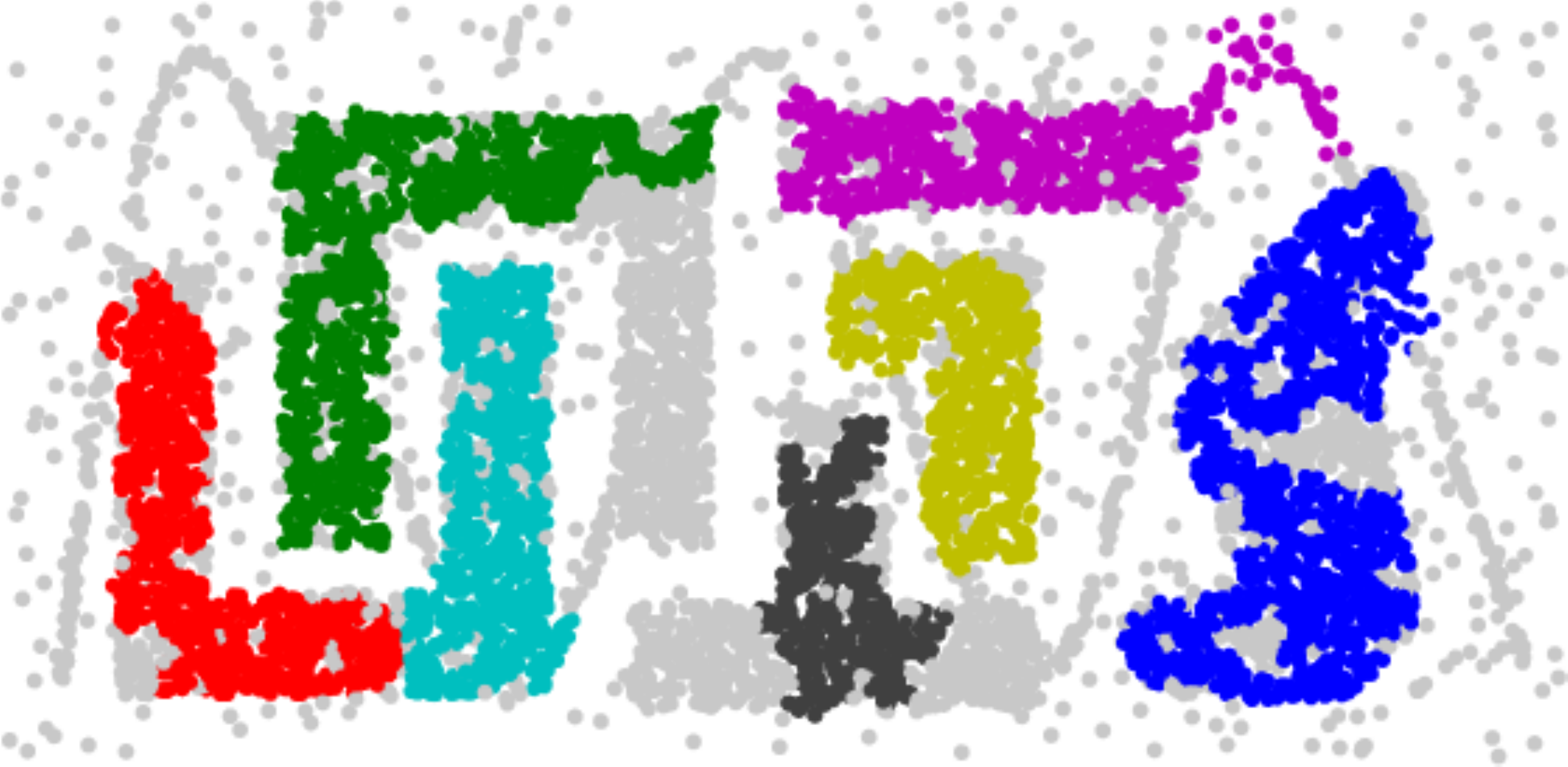}}\tabularnewline
(e) & (f)\tabularnewline
\end{tabular}
\par\end{centering}

\protect\caption{\label{fig:clustering_data_results_DS4}Clustering results of a dataset
taken from \cite{karypis_chameleon:_1999}. (a) 1 cluster; (b) 2 clusters;
(c) 4 clusters; (d) 5 clusters; (e) 6 clusters; (f) 7 clusters. }
\end{figure}

Using the proposed algorithm, we cluster a dataset taken from \cite{karypis_chameleon:_1999}.
Fig. \ref{fig:clustering_data_results_DS4} shows some results of
different number of detected clusters. Outliers are shown in light
gray color and data points in different clusters are shown in different
colors. 

As the Fig. \ref{fig:clustering_data_results_DS4} shows, the proposed
algorithm cannot only classify outliers and normal data points but
also find clusters in the data points. As more and more edges are
removed from the MKNN graph, the number of clusters increases.

Next we show how to determine the true number of clusters. Table \ref{tab:number_removed_edges_in_DS4}
shows the number of removed edges and the number of detected clusters
of this dataset. 

\begin{table}[htbp]
\protect\caption{\label{tab:number_removed_edges_in_DS4}Percentage of the Removed
Edges and the Number of Detected Clusters}

\centering{}%
\begin{tabular}{ccccccc}
\toprule 
removed edges & 2.6\% & 2.7\% & 2.8\% & 3.5\% & 6\% & 33.3\%\tabularnewline
\midrule 
number of clusters & 2 & 3 & 4 & 5 & 6 & 7\tabularnewline
\bottomrule
\addlinespace
\end{tabular}
\end{table}

As the result shows, removing a small amount of edges is enough to
find correct clusters in the data. One has to remove a large amount
of edges to break a genuine cluster into smaller components. We can
simply define a threshold and stop the iteration if the number of
clusters does not increase any more. 

\begin{figure}[tbh]
\begin{centering}
\textsf{}%
\begin{tabular}{ccc}
\textsf{\includegraphics[width=2.5cm]{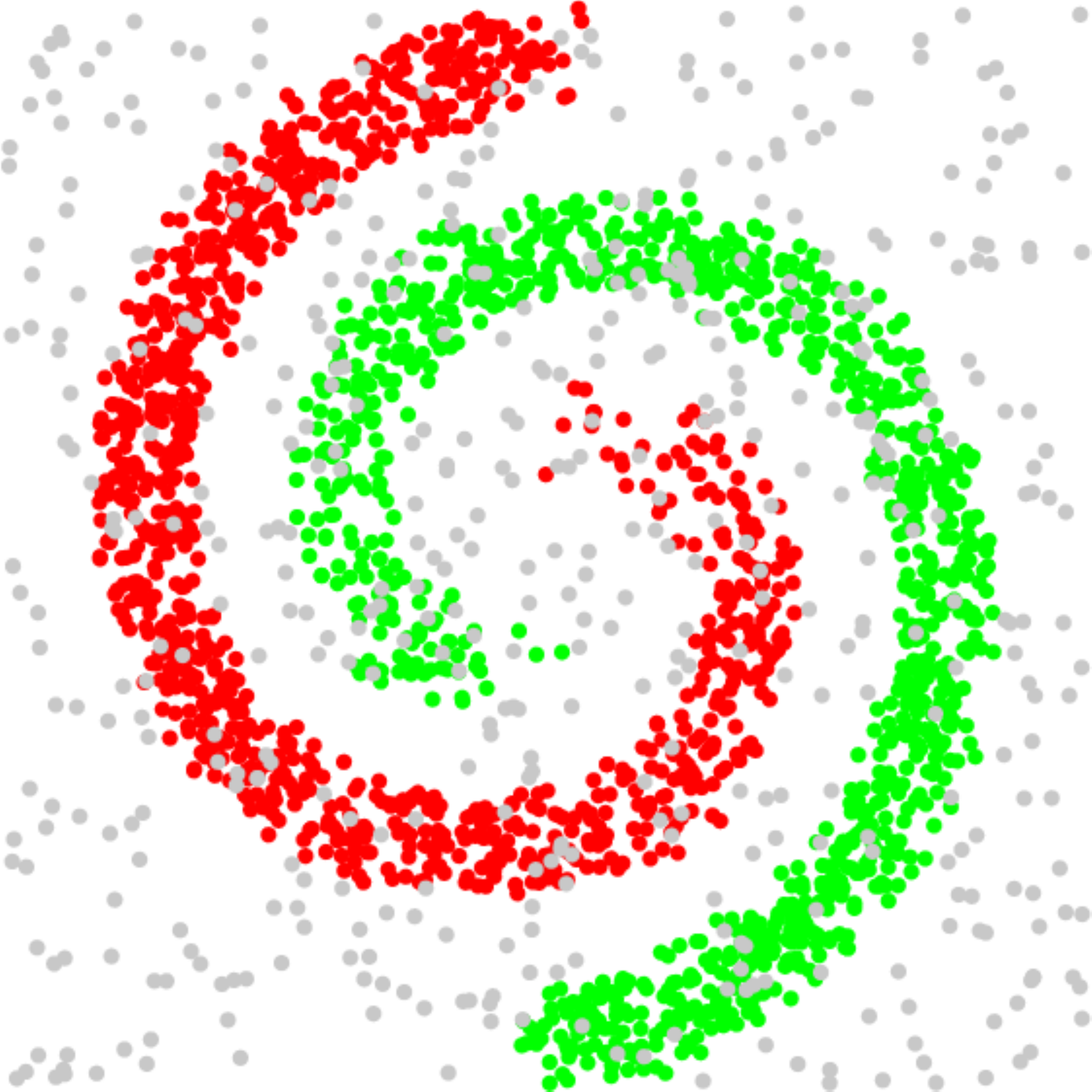}} & \includegraphics[width=2.5cm]{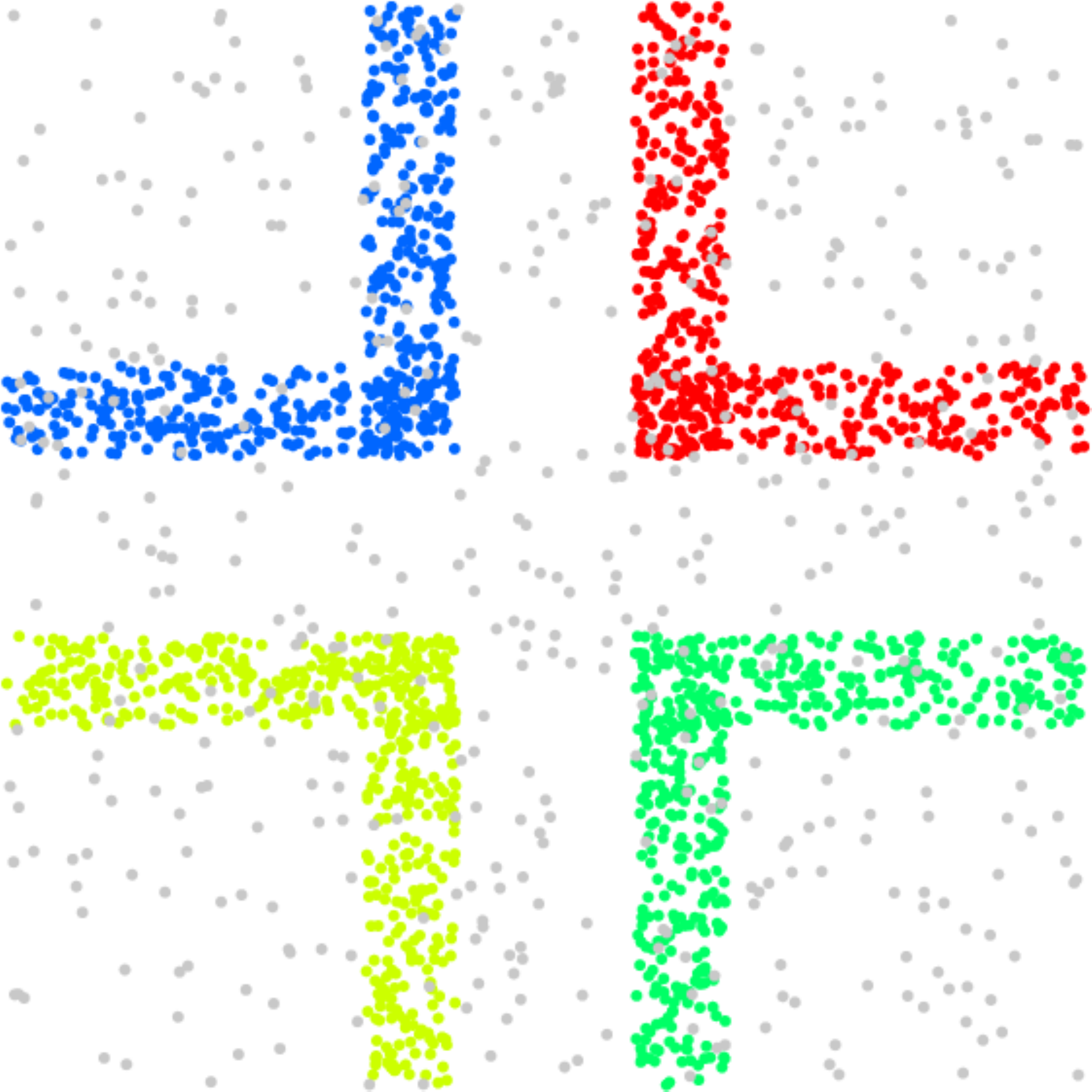} & \includegraphics[width=2.5cm]{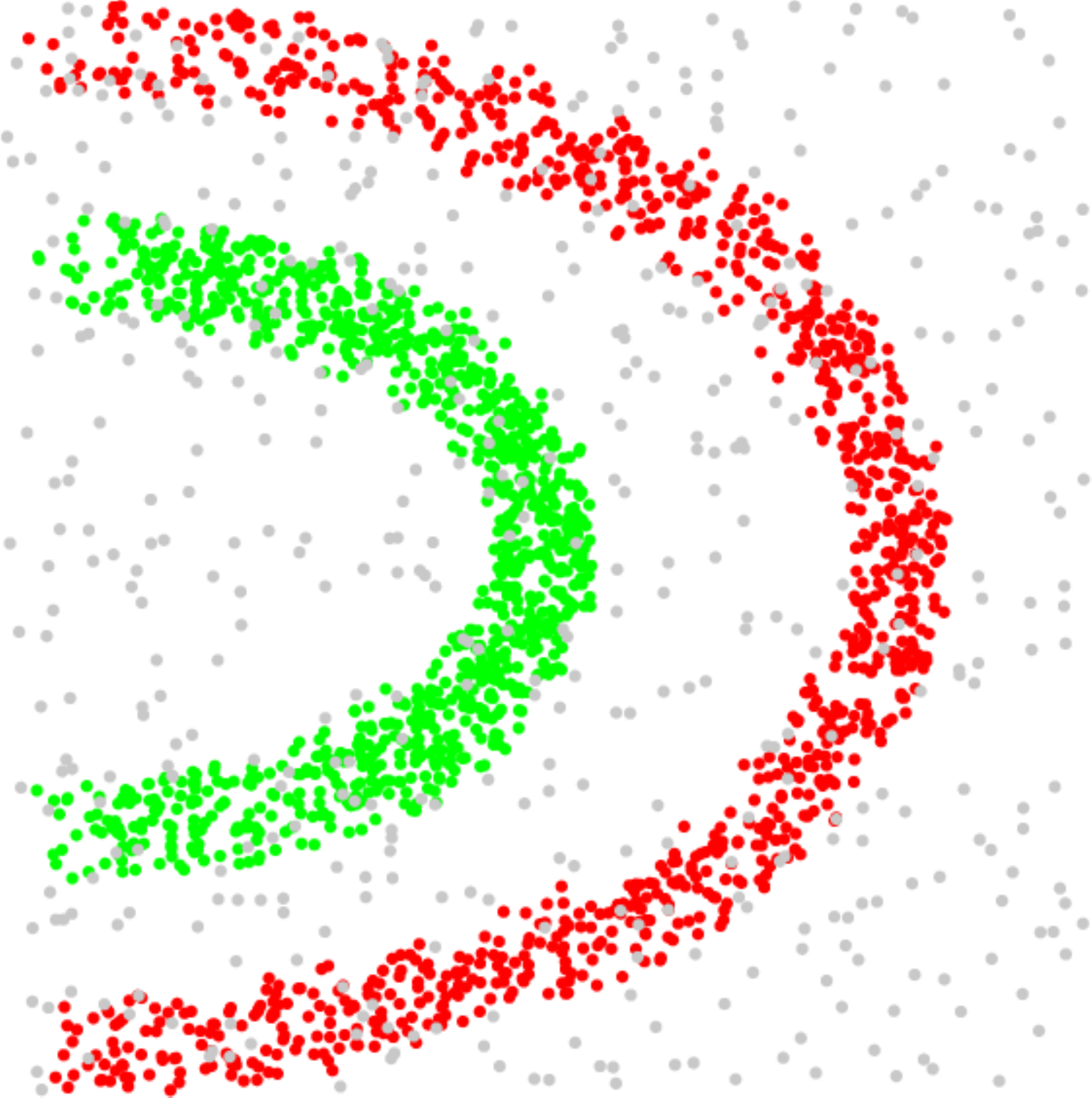}\tabularnewline
(a) & (b) & (c)\tabularnewline
 &  & \tabularnewline
\includegraphics[width=2.5cm]{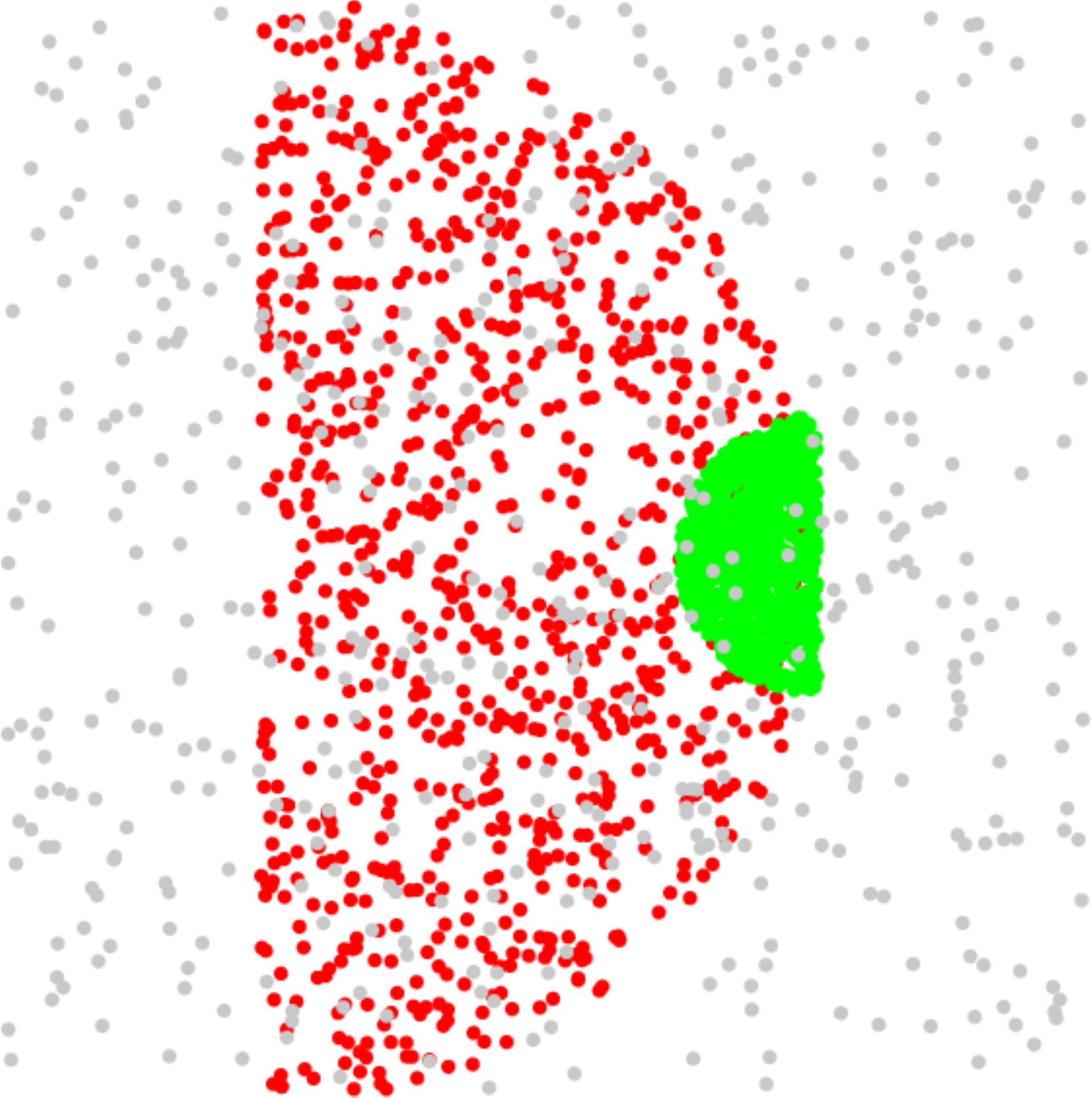} & \includegraphics[width=2.5cm]{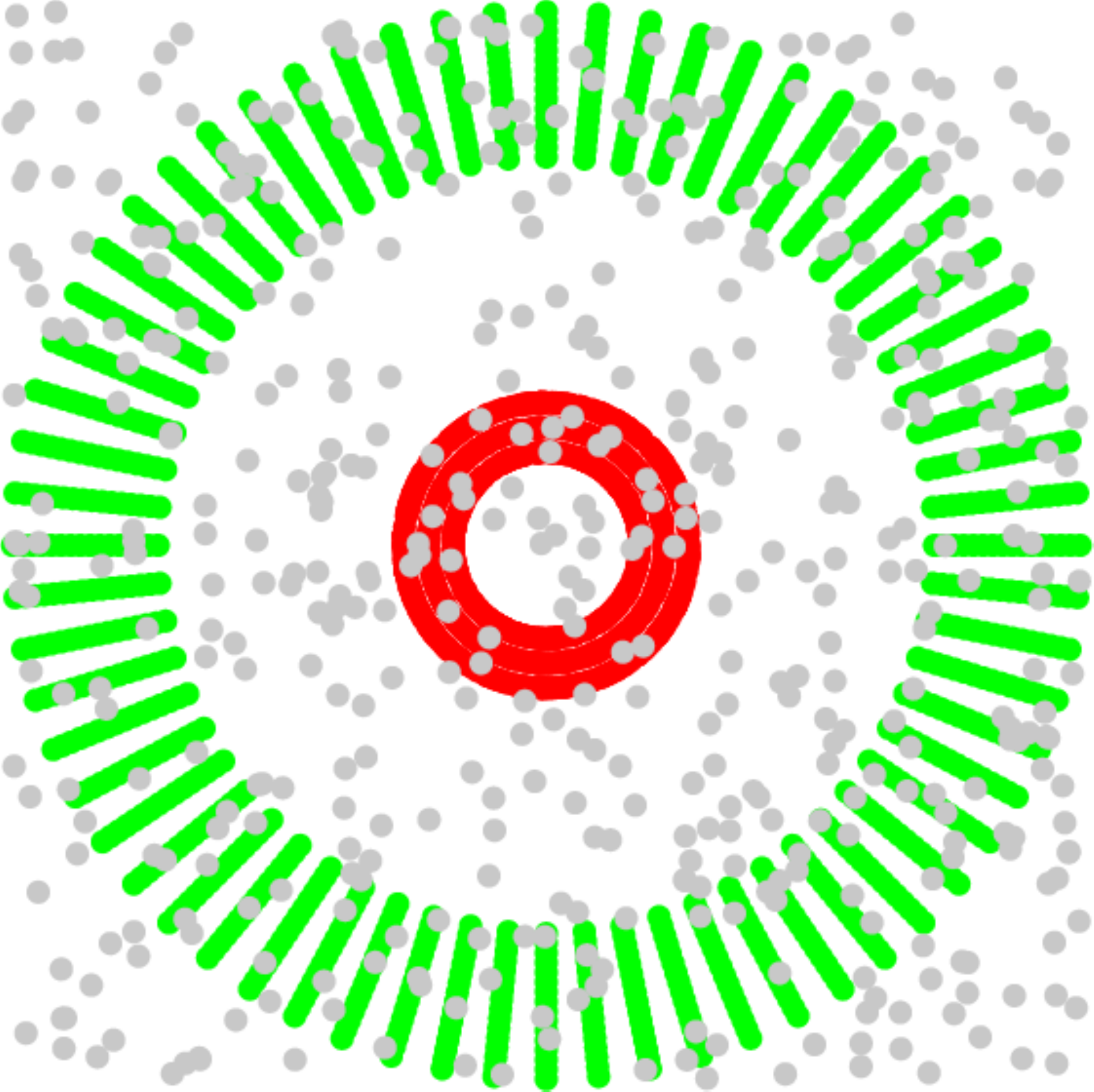} & \includegraphics[width=2.5cm]{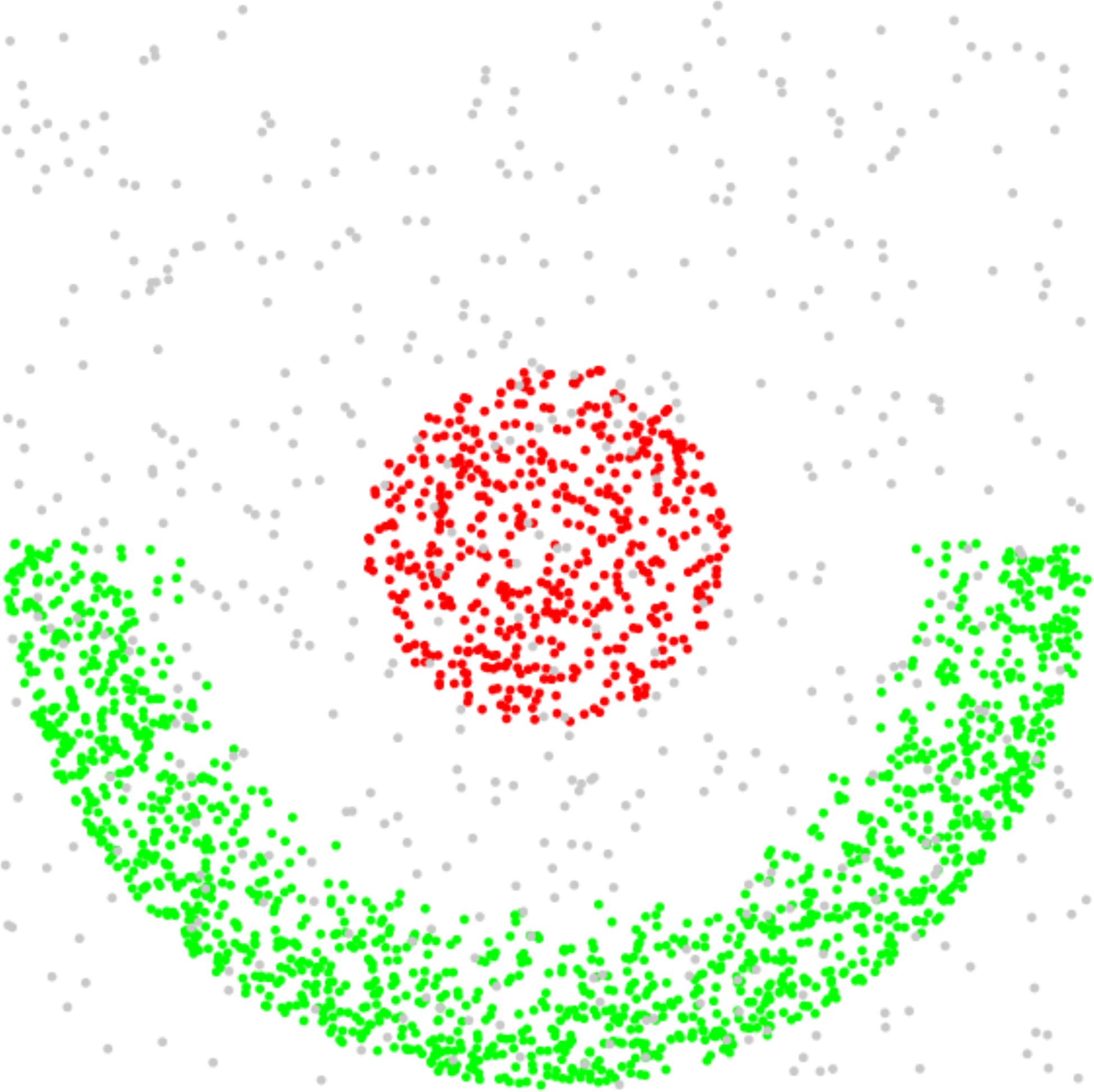}\tabularnewline
(d) & (e) & (f)\tabularnewline
\end{tabular}
\par\end{centering}

\protect\caption{\label{fig:clustering_data}Synthetic datasets for clustering}
\end{figure}

To illustrate the performance of the proposed clustering algorithm,
we use synthetic data that are both noisy and challenging. Fig. \ref{fig:clustering_data}
shows the test datasets. We used tools from \cite{_6_????} to generate
the normal data points and added random data points as noise. 

In our experiments, we use the Euclidean distance function. The number
of nearest neighbors is 30. At each iteration step, we remove 0.1\%
of total number of edges according to their outlier scores. A large
connected component is a component whose size is larger than 5\% of
the total number of nodes. The clustering termination threshold is
set as 10\% of the total number of edges. 

We compare the proposed clustering algorithm with the k-means\cite{theodoridis_pattern_2008},
the average-linkage (a-link)\cite{theodoridis_pattern_2008}, the
normalized cuts (N-Cuts)\cite{shi_normalized_2000} and the graph
degree linkage (GDL)\cite{zhang_graph_2012} clustering algorithms.
Since the competing algorithms cannot detect the number of clusters,
we use the value from the ground truth. Table \ref{tab:clustering_results_nmi}
shows the NMI scores of the proposed algorithm and the competing algorithms. 

\begin{table}[htbp]
\protect\caption{\label{tab:clustering_results_nmi}Clustering of Noisy Data Results}

\centering{}%
\begin{tabular}{cccccc}
\toprule 
dataset & k-means & a-link & N-Cuts & GDL & proposed\tabularnewline
\midrule
\addlinespace
(a) & 0.031 & 0.099 & 0.053 & 0.650 & \textbf{0.672}\tabularnewline
\addlinespace
(b) & 0.743 & 0.743 & 0.743 & 0.743 & \textbf{0.848}\tabularnewline
\addlinespace
(c) & 0 & 0.004 & 0.559 & 0.654 & \textbf{0.755}\tabularnewline
\addlinespace
(d) & 0.208 & 0.161 & 0.367 & 0.553 & \textbf{0.619}\tabularnewline
\addlinespace
(e) & 0.001 & 0.133 & 0.680 & 0.701 & \textbf{0.744}\tabularnewline
\addlinespace
(f) & 0.001 & 0.162 & 0.627 & 0.612 & \textbf{0.714}\tabularnewline
\bottomrule
\addlinespace
\end{tabular}
\end{table}

The results show that the k-means and the average linkage clustering
algorithms fail on complex-shaped clusters. GDL and the proposed algorithms
are all graph-based clustering algorithms. They are able to find clusters
with arbitrary shapes. From the NMI scores, the proposed algorithm
is clearly superior to the competing clustering algorithms.

\section{Conclusions}

In real-world graphs, in particular social network graphs, there are
edges generated by scammers, malicious programs or mistakenly by normal
users and the system. Detecting these outlier edges and removing them
will not only improve the efficiency of graph mining and analytics,
but also help identify harmful entities. In this article, we introduce
outlier edge detection algorithms based on two random graph generation
models. We define four schemes that represent relationships of two
nodes and the groups of their neighboring nodes. We combine the schemes
with the two random graph generation models and investigate the proposed
algorithms theoretically. We tested the proposed outlier edge detection
algorithms by experiments on real-world graphs. The experimental results
show that our proposed algorithms can effectively identify the injected
edges in real-world graphs. We compared the performance of our proposed
algorithms with other outlier edge detection algorithms. The proposed
algorithms, especially the algorithm based on the PA model, give consistently
good results regardless of the test graph data. We also evaluated
the changes of graph properties caused by the removal of the detected
outlier edges. The experimental results show an increase in both the
clustering coefficients and the increase of the distance between the
nodes in the graph. This is coherent with the theoretical predictions. 

Further more, we demonstrate the potential of the outlier edge detection
using three different applications. When used with the graph clustering
algorithms, removing outlier edges from the graph not only improves
the clustering accuracy but also reduces the computational time. This
indicates that the proposed algorithms are powerful preprocessing
tools for graph mining. When used for detecting outlier nodes in social
network graphs, we can successfully find outlier nodes whose behavior
deviates dramatically from that of normal nodes. We also present a
clustering algorithm that is based on the edge outlier scores. The
clustering algorithm can efficiently find true data clusters by excluding
noises from the data. 

Outlier edge detection has great potentials in numerous Big Data applications.
In the future, we will apply the proposed outlier edge detection algorithms
in applications in other fields, for example computer vision and content-based
multimedia retrieval in the Big Visual Data. We observed that nodes
and edges outside edge-ego-network also contain valuable information
in outlier detection. However, using this information dramatically
increases the computational cost. We will work on fast algorithms
that can efficiently use the structural information of the whole graph.

\appendices{}

\section*{Proof of Theorem \ref{thm:scheme2_4_symmetric}}
\begin{prop}
\label{prop:distributive_law}$\alpha\left(S\cup T,R\right)=\alpha\left(S,R\right)+\alpha\left(T,R\right)$
if $S\cap T=\emptyset$.\end{prop}
\begin{IEEEproof}
Let $A$ be the adjacency matrix of an unweighted and undirected graph
$G$. We have $\alpha(S,T)=\sum_{i\in S}\sum_{j\in T}A_{ij}$. Given
$S\cap T=\emptyset$, 

\begin{align*}
\alpha\left(S\cup T,R\right) & =\sum_{i\in S\cup T}\sum_{j\in R}A_{ij}\\
 & =\sum_{i\in S}\sum_{j\in R}A_{ij}+\sum_{i\in T}\sum_{j\in R}A_{ij}\\
 & =\alpha(S,R)+\alpha(T,R)
\end{align*}

\end{IEEEproof}
Next we prove Theorem \ref{thm:scheme2_4_symmetric}. 
\begin{IEEEproof}
For scheme 4, $P_{a,b}^{(4)}=S_{a\backslash b}$, $R_{a,b}^{(4)}=S_{b\backslash a}$,
$P_{b,a}^{(4)}=S_{b\backslash a}$ and $R_{b,a}^{(4)}=S_{a\backslash b}$.
Using Theorem \ref{thm:alpha_epsilon_symmetric}, we can easily get
$\alpha\left(P_{a,b}^{(4)},R_{a,b}^{(4)}\right)=\alpha\left(P_{b,a}^{(4)},R_{b,a}^{(4)}\right)$. 

To prove Theorem \ref{thm:scheme2_4_symmetric} for scheme 2, we divide
the nodes in edge-ego-network $G_{\overline{ab}}$ into five mutually
exclusive sets:
\begin{itemize}
\item $V_{1}=\left\{ x\vert x\in N_{a}\ \mathrm{and}\ x\notin S_{b}\right\} $; 
\item $V_{2}=\left\{ x\vert x\in N_{b}\ \mathrm{and}\ x\notin S_{a}\right\} $;
\item $V_{3}=\left\{ x\vert x\in N_{a}\ \mathrm{and}\ x\in N_{b}\right\} $;
\item $V_{4}=\left\{ a\right\} $;
\item $V_{5}=\left\{ b\right\} $.
\end{itemize}
From the definition, we have

\begin{eqnarray*}
P_{a,b}^{(2)} &  & =N_{a\backslash b}=V_{1}\cup V_{3},\\
R_{a,b}^{(2)} &  & =S_{b\backslash a}=V_{2}\cup V_{3}\cup V_{5},\\
P_{b,a}^{(2)} &  & =N_{b\backslash a}=V_{2}\cup V_{3},\\
R_{b,a}^{(2)} &  & =S_{a\backslash b}=V_{1}\cup V_{3}\cup V_{4}.
\end{eqnarray*}

Using the definition of $\alpha(S,T)$ and Proposition \ref{prop:distributive_law},
we get

\begin{eqnarray}
\alpha\left(P_{a,b}^{(2)},R_{a,b}^{(2)}\right) & = & \alpha\left(V_{1}\cup V_{3},V_{2}\cup V_{3}\cup V_{5}\right)\label{eq:scheme2_ab}\\
 & = & \alpha\left(V_{1},V_{2}\right)+\alpha\left(V_{1},V_{3}\right)+\alpha\left(V_{1},V_{5}\right)\nonumber \\
 &  & +\alpha\left(V_{3},V_{2}\right)+\alpha\left(V_{3},V_{3}\right)+\alpha\left(V_{3},V_{5}\right)\nonumber 
\end{eqnarray}

and 
\begin{eqnarray}
\alpha\left(P_{b,a}^{(2)},R_{b,a}^{(2)}\right) & = & \alpha\left(V_{2}\cup V_{3},V_{1}\cup V_{3}\cup V_{4}\right)\label{eq:scheme2_ba}\\
 & = & \alpha\left(V_{2},V_{1}\right)+\alpha\left(V_{2},V_{3}\right)+\alpha\left(V_{2},V_{4}\right)\nonumber \\
 &  & +\alpha\left(V_{3},V_{1}\right)+\alpha\left(V_{3},V_{3}\right)+\alpha\left(V_{3},V_{4}\right)\nonumber 
\end{eqnarray}

Taking the fact that $\alpha(V_{1}\cap V_{5})=0$, $\alpha(V_{2}\cap V_{4})=0$,
and $\alpha(V_{3},V_{4})=\alpha(V_{3},V_{5})$, the right hand side
of Eqs. \ref{eq:scheme2_ab} and \ref{eq:scheme2_ba} are equal. Thus
$\alpha\left(P_{a,b}^{(2)},R_{a,b}^{(2)}\right)=\alpha\left(P_{b,a}^{(2)},R_{b,a}^{(2)}\right)$. 
\end{IEEEproof}

\bibliographystyle{IEEEtran}
\bibliography{clean}

\begin{IEEEbiography}[{\fbox{\begin{minipage}[t][1.25in]{1in}%
\includegraphics[clip,width=1in,height=1.25in]{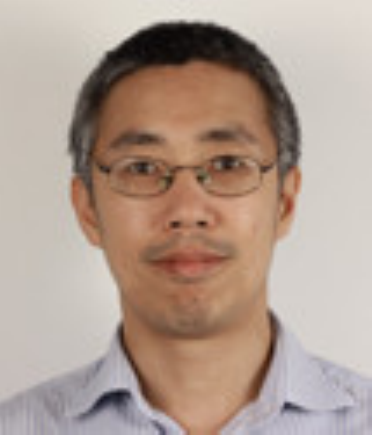}%
\end{minipage}}}]{Honglei Zhang}
is a PhD student and a researcher in the Department of Signal Processing
at Tampere University of Technology. He received his Bachelor and
Master degree in Electrical Engineering from Harbin Institute of Technology
in 1994 and 1996 in China, respectively. He worked as a software engineer
in Founder Co. in China from 1996 to 1999. He had been working as
a software engineer and a software system architect in Nokia Oy Finland
for 14 years. He has 6 scientific publications and 2 patents. His
current research interests include computer vision, pattern recognition,
data mining and graph algorithms. More about of his research can be
found from \url{http://www.cs.tut.fi/~zhangh}.
\end{IEEEbiography}

\begin{IEEEbiography}[{\fbox{\begin{minipage}[t][1.25in]{1in}%
\includegraphics[clip,width=1in,height=1.25in]{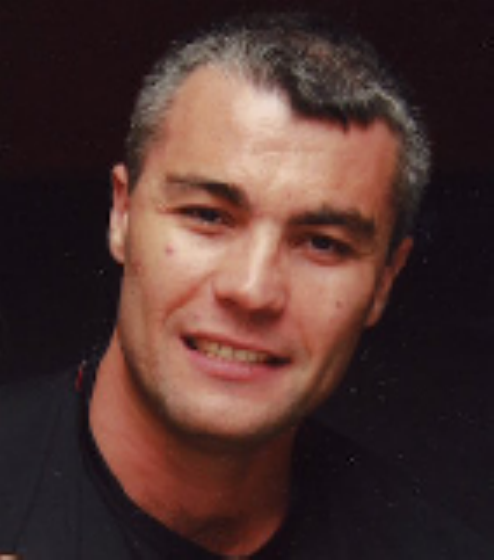}%
\end{minipage}}}]{Serkan Kiranyaz}
was born in Turkey, 1972. He received his BS degree in Electrical
and Electronics Department at Bilkent University, Ankara, Turkey,
in 1994 and MS degree in Signal and Video Processing from the same
University, in 1996. He worked as a Senior Researcher in Nokia Research
Center and later in Nokia Mobile Phones, Tampere, Finland. He received
his PhD degree in 2005 and his Docency at 2007 from Tampere University
of Technology, respectively. He is currently a Professor in the Department
of Electrical Engineering at Qatar University. Prof. Kiranyaz published
2 books, more than 30 journal papers in several IEEE Transactions
and some other high impact journals and 70+ papers in international
conferences. His recent publication has been nominated for the Best
Paper Award in IEEE ICIP\textquoteright 13 conference. Another publication
won the IBM Best Paper Award in ICPR\textquoteright 14.
\end{IEEEbiography}

\vfill{}

\pagebreak{}
\begin{IEEEbiography}[{\fbox{\begin{minipage}[t][1.25in]{1in}%
\includegraphics[clip,width=1in,height=1.25in]{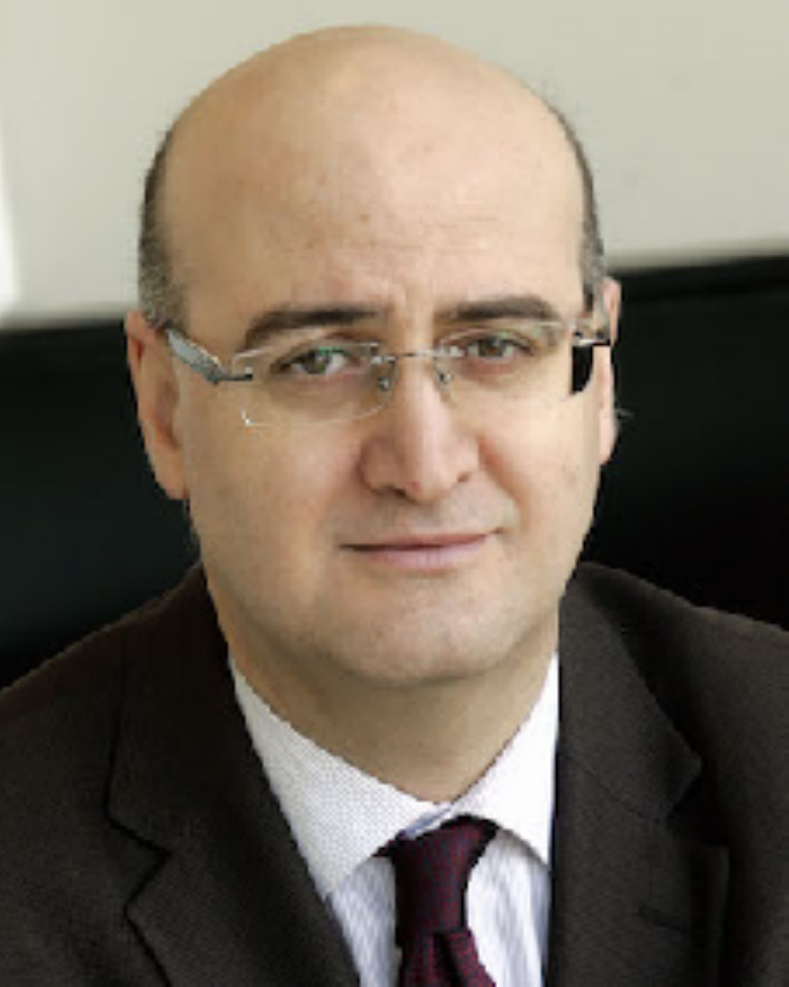}%
\end{minipage}}}]{Moncef Gabbouj}
received his BS degree in electrical engineering in 1985 from Oklahoma
State University, and his MS and PhD degrees in electrical engineering
from Purdue University, in 1986 and 1989, respectively. Dr. Gabbouj
is currently Academy of Finland Professor and holds a permanent position
of Professor of Signal Processing at the Department of Signal Processing,
Tampere University of Technology. His research interests include multimedia
content-based analysis, indexing and retrieval, machine learning,
nonlinear signal and image processing and analysis, voice conversion,
and video processing and coding. Dr. Gabbouj is a Fellow of the IEEE
and member of the Finnish Academy of Science and Letters. He is the
past Chairman of the IEEE CAS TC on DSP and committee member of the
IEEE Fourier Award for Signal Processing. He served as Distinguished
Lecturer for the IEEE CASS. He served as associate editor and guest
editor of many IEEE, and international journals. \end{IEEEbiography}

\end{document}